\pdfoutput=1

\documentclass[10pt]{article}

\usepackage{listings}    % for snippets of code using \begin{lstlisting}.
\usepackage{graphicx}    % for figures, eg, \includegraphics
\usepackage{subcaption}  % for subfigures, ie figures with internal figures
\usepackage{algorithm}   % for displaying algorithm in pseudo-code, using \begin{algorithm}
\usepackage{algorithmic} % and \begin{algorithmic}
\usepackage{booktabs}    % for better tables, eg using \toprule
\usepackage{hyperref}    % to handle URL links with \url
\usepackage{xurl}        % to handle very long URLs inside \url{} (done automatically, nothing to set manually)
\usepackage[htt]{hyphenat} % for line breaks in texttt, using \-
\usepackage{microtype} % ligatures are extremely annoying, especially when copy&paste text from PDF
\DisableLigatures{}    % https://tex.stackexchange.com/questions/439651/how-do-i-disable-ligatures
\usepackage{adjustbox} % for resizing tables eg with "\begin{adjustbox}{width=.95\textwidth,center}".
\usepackage{colortbl}  % for colors in table rows, eg using \rowcolor{<color>}

\usepackage{caption}
\usepackage{setspace}
\usepackage{multirow}
\usepackage{enumerate}
\usepackage{pdflscape}
\usepackage{pgfplots} % bar lib

\usepackage{xcolor}
\definecolor{codegreen}{rgb}{0.25,0.5,0.35}
\definecolor{codegray}{rgb}{0.5,0.5,0.5}
\definecolor{codepurple}{rgb}{0.6,0,0}
\definecolor{backcolour}{rgb}{0.95,0.95,0.92}
\definecolor{colorstring}{rgb}{0.5,0,0.35}
\definecolor{rltred}{rgb}{0.5,0,0}
\definecolor{rltgreen}{rgb}{0,0.5,0}
\definecolor{rltblue}{rgb}{0,0,0.5}
\definecolor{DarkGreen}{rgb}{0.00,0.60,0.00}
\definecolor{ScarletRed}{rgb}{0.80,0.00,0.00}
\definecolor{blizzardblue}{rgb}{0.67, 0.9, 0.93}
\definecolor{green-yellow}{rgb}{0.68, 1.0, 0.18}
\definecolor{dkgreen}{rgb}{0,0.6,0}
\definecolor{gray}{rgb}{0.5,0.5,0.5}
\definecolor{mauve}{rgb}{0.58,0,0.82}
\definecolor{lightgrey}{rgb}{0.90,0.90,0.90}
\definecolor{grey}{gray}{0.75}
\definecolor{light-gray}{gray}{0.80}

\lstdefinestyle{mystyle}{
    escapechar=©, %  use ©\label{}© when needing \label pointing to line numbers
	backgroundcolor=\color{backcolour},
    basicstyle=\scriptsize\ttfamily,
   	identifierstyle=\footnotesize\ttfamily,
	commentstyle=\color{codegreen},
	keywordstyle=\color{colorstring}\bfseries,
	numberstyle=\ttfamily\color{codegray},
	stringstyle=\ttfamily\color{DarkGreen},
	breakatwhitespace=false,
	breaklines=true,
	captionpos=b,
	keepspaces=true,
	numbers=left, % possible values are (none, left, right)
	numbersep=2pt,
	showspaces=false,
	showstringspaces=false,
	showtabs=false,
	tabsize=2,
	inputencoding=utf8,
    extendedchars=true,
    literate={ø}{{\o}}1 {æ}{{\ae}}1 {å}{{\aa}}1
}
\usepackage{xspace}
\newcommand{\evo}{{\sc EvoMaster}\xspace}

\usepackage{boxedminipage}
\newenvironment{result}%
{\smallskip
	\noindent
	\let\emph=\textbf
	\begin{boxedminipage}{\columnwidth}\begin{center}\em}%
		{\end{center}\end{boxedminipage}%
}

\usepackage{ifthen}
\newboolean{showcomments}
\setboolean{showcomments}{true} % comment this line to deactivate comments

\ifthenelse{\boolean{showcomments}}{
	\newcommand{\nbc}[3]{
		{\colorbox{#3}{\bfseries\sffamily\scriptsize\textcolor{white}{#1}}}
		{\textcolor{#3}{\sf\small$\langle$\textit{#2}$\rangle$}}}
	
}{
	\newcommand{\nbc}[3]{}

}

\usepackage{authblk}                % for \affil
\usepackage[square,numbers]{natbib} % needed for the references, when using ACM-Reference-Format outside an
\usepackage{amsmath}                % for math operations
\usepackage{amssymb}                % for symbols like \checkmark

\usepackage{geometry}
\title{
Enhancing REST API Fuzzing with Access Policy Violation Checks and Injection Attacks
}

\author[1,2]{Omur Sahin}
\author[3]{Man Zhang}
\author[2,4]{Andrea Arcuri}

\affil[1]{Erciyes University, Türkiye}
\affil[2]{Kristiania University College, Norway}
\affil[3]{Beihang University, China}
\affil[4]{Oslo Metropolitan University, Norway}

\date{}

\begin{document}

\maketitle

\begin{abstract}
Due to their widespread use in industry, several techniques have been proposed in the literature to fuzz REST APIs.
Existing fuzzers for REST APIs have been focusing on detecting crashes (e.g., 500 HTTP server error status code).
However, security vulnerabilities can have major drastic consequences on existing cloud infrastructures.

In this paper, we propose a series of novel automated oracles aimed at detecting violations of access policies in REST APIs,
as well as executing traditional attacks such as SQL Injection and XSS.
These novel automated oracles can be integrated into existing fuzzers, in which, once the fuzzing session is completed,
a ``security testing'' phase is executed to verify these oracles.
When a security fault is detected, as output our technique is able to general executable test cases in different formats,
like Java, Kotlin, Python and JavaScript test suites.

Our novel techniques are integrated as an extension of \evo, a state-of-the-art open-source fuzzer for REST APIs.
Experiments are carried out on 9 artificial examples, 8 vulnerable-by-design REST APIs with \emph{black-box} testing, and 36 REST APIs from the WFD corpus with \emph{white-box} testing, for a total of 52 distinct APIs.
Results show that our novel oracles and their automated integration in a fuzzing process can lead to detect security issues in several
of these APIs.
\end{abstract}


%%%%%%%%%%%%%%%%%%%%%%%%%%%%%%%%%%%%%%%%%%%%%%%%%%%%%%%%%%%%%%%%%%%%%%%%%%%%%%%%%%%%%%%%%%%%%%%%%%%%%%%%%%%%%%%%%%%%%%%
\section{Introduction}

The Open Worldwide Application Security Project (OWASP) keeps track of the most common security vulnerabilities in web
applications.\footnote{\label{owasp-web}\url{https://owasp.org/www-project-top-ten/}}
In their 2023-report targeted at web APIs,\footnote{\label{owasp-api}\url{https://owasp.org/www-project-api-security/}}
the most common vulnerability is \emph{Broken Object Level Authorization} (BOLA).
A similar vulnerability, called \emph{Broken Function Level Authorization} (BFLA), ranks fifth.
Wrong authorization checks in the APIs can lead users to access resources they are not supposed to access (BOLA),
or do operations they do not have rights to (BFLA).

Based on the ``State of API Security 2026'' made
by \emph{42crunch},\footnote{\label{foot:42crunch}\url{https://42crunch.com/state-of-api-security-2026-report/}}
which is based on 200 real-world API vulnerabilities and exploits discussed in the \emph{APISecurity.io} newsletter coverage between 2024 and 2025,
``\emph{the most frequent API vulnerability across all categories in 2025 was {\bf Missing Authentication}.
This occurs when an API fails to enforce authentication for sensitive operations, allowing anonymous users
to access data or perform actions that should be restricted to authenticated users}''.\footref{foot:42crunch}

As of 2024, it is estimated\footnote{\url{https://edgedelta.com/company/blog/how-many-companies-use-cloud-computing}}
that 94\% of companies worldwide use cloud computing, with costs measured in the hundreds of billions of euros.
As such, reliability and correctness of these systems is of paramount importance in modern society.

Currently, REST APIs are the most common type of APIs on the web.
As testing the functionalities of APIs can be expensive and time consuming, a lot of research has been carried out to
automatically generate test cases for them~\cite{golmohammadi2023testing}.
However, to deal with authorization-related faults, a fuzzer should be able to handle the authentication of different users
with different roles and access rights.
Unfortunately, many fuzzers in the literature are unable to handle authentication information during the fuzzing process~\cite{sahin_2025_wfc}.
For the few fuzzers that can handle authentication of different users, they are unable to automatically detect BOLA and BFLA related
vulnerabilities.

To fill this gap in the literature,
in this paper  we define seven novel automated oracles aimed at detecting faults related to misconfigured authorization checks.
Furthermore, as the verification of authorization checks is only one aspect of security testing, we also integrated in our approach the evaluation of SQL Injection and XSS attacks.

We integrated our novel oracles in a state-of-the-art fuzzer for REST APIs, namely \evo~\cite{arcuri2025tool}.
Experiments on eight artificial APIs widely used in security research, as well as on 36 APIs from the WFD corpus~\cite{sahin_2025_wfc}, show that our novel automated oracles can be used to automatically detect several new security vulnerabilities that cannot be detected with
other current approaches in the literature.

Our approach is integrated and evaluated in \evo, but any other state-of-the-art fuzzer
that supports authentication of users (e.g., via authorization headers, or dynamic login tokens)
could had been used for our study.
We chose to use \evo as starting point for our implementation, as developing a robust fuzzer from scratch is a major
engineering effort~\cite{arcuri2023building}.
We chose \evo because it is among the oldest~\cite{arcuri2017restful} and mature fuzzer still in active development~\cite{arcuri2025tool},
giving among the best results in large tool comparisons~\cite{Kim2022Rest,zhang2023open,sartaj2024restapitestingdevops}.
Although \evo is focused on \emph{white-box} testing, its \emph{black-box} mode was able to achieve second position in the SBFT'26 black-box fuzzer competition~\cite{arcuri2026sbft}.
As it is a mature tool with extensive documentation, it is used in several enterprises worldwide, such as
Meituan in China~\cite{zhang2023rpc,zhang2024seeding} and Volkswagen in Germany~\cite{poth2025technology,icst2025vw}.
\evo has been downloaded thousands of times~\cite{arcuri2025tool}.

This paper is an extension of conference paper~\cite{arcuri2025fuzzing}.
%In that study, only 3 oracles were presented and analyzed, on a much smaller empirical study (13 APIs).
Compared with the conference version, we introduced six additional novel oracles (9 \textit{vs.} 3) and conducted empirical studies on a larger set of APIs (52 \textit{vs.} 13).
Besides the novel security oracles presented in this paper, \evo is also able to detect Server-Side Request Forgery (SSRF) vulnerabilities~\cite{seran2026sbft}.
However, the details of how it is done have been presented and analyzed in a different previous publication~\cite{seran2026sbft}, and they are not going to be discussed here.

The paper is organized as follows.
Section~\ref{sec:related} discusses related work.
Our novel security oracles are presented in Section~\ref{sec:oracles}.
How the use of these oracles can be integrated into a fuzzer is discussed in Section~\ref{sec:integration}.
Section~\ref{sec:empirical} shows our empirical study.
Threats to validity follows in Section~\ref{sec:threats}.
Finally, Section~\ref{sec:conclusions} concludes the paper.

%%%%%%%%%%%%%%%%%%%%%%%%%%%%%%%%%%%%%%%%%%%%%%%%%%%%%%%%%%%%%%%%%%%%%%%%%%%%%%%%%%%%%%%%%%%%%%%%%%%%%%%%%%%%%%%%%%%%%%%
\section{Related Work}
\label{sec:related}

There is a large body of research related to fuzz REST APIs~\cite{golmohammadi2023testing}.
Several fuzzers have been proposed in the literature, including, in alphabetic order:
APIRL~\cite{foley2025apirl},
APIF~\cite{wang2024beyond},
ARAT-RL\cite{kim2023adaptive},
AutoRestTest~\cite{kim2025autoresttest},
bBOXRT~\cite{laranjeiro2021black},
DeepREST~\cite{corradini2024deeprest},
\evo~\cite{arcuri2025tool},
LlamaRestTest~\cite{kim2025llamaresttest},
MINER~\cite{lyu2023miner},
Morest~\cite{liu2022icse},
Nautilus~\cite{deng2023nautilus},
ResTest~\cite{martinLopez2021Restest},
RestCT~\cite{wu2022icse},
RESTler~\cite{restlerICSE2019},
RestTestGen~\cite{viglianisi2020resttestgen},
Sche\-ma\-thesis~\cite{hatfield2022deriving},
VoAPI2~\cite{du2024vulnerability}
and
WuppieFuzz~\cite{rooijakkers2025wuppiefuzz}.
These fuzzers aim at achieving high coverage (e.g., in terms of schema coverage and code coverage),
and detecting faults.
The most common type of oracle is detecting server errors~\cite{marculescu2022faults}, usually represented with a 500 HTTP status code~\cite{golmohammadi2023testing}.
Other types of oracles include checking for schema mismatches (i.e., the test API is returning data that is not valid according to
the API's schema), and robustness properties (i.e., sending wrong data on purpose, and then check whether the API correctly flags them as user errors).
In large empirical comparisons~\cite{Kim2022Rest,zhang2023open}, fuzzers for REST APIs are typically compared on code coverage
and on the number of detected 500 status codes.

When it comes to verify security properties in REST APIs, some initial work on this topic exists.
One of the earlier approaches was in RESTler~\cite{atlidakis2020checking}, in which 4 security rules were defined.
These included checking the semantics of HTTP operations
(e.g., ``\emph{A resource that has been deleted must no longer be accessible}''~\cite{atlidakis2020checking})
as well as checking user-namespace properties (e.g.,
``\emph{A resource created in a user namespace must not be accessible from another user namespace}''~\cite{atlidakis2020checking}).
However, this latter rule might be too restrictive, and not considering that there can be complex access control policies in
place where some users might have access to some resources of other users (e.g., a typical case would be administrator users).

The work presented in Nautilus aims at detecting ``\emph{vulnerabilities caused by improper user input handling}''~\cite{deng2023nautilus}.
However, the presented approach is not fully automated, as it requires the testers to manually annotate the OpenAPI schemas
with extra information needed by Nautilus to detect vulnerabilities.
VoAPI2~\cite{du2024vulnerability} does different kinds of security attacks, including  SSRF, XSS, and Command Injection testing.
Different techniques are used to identify subsets of the API operations and parameters that are more likely to be affected by
these kinds of security issues.
Similarly, APIF~\cite{wang2024beyond} does different kinds of security attacks based
on SecLists,\footnote{\url{https://github.com/danielmiessler/SecLists}}
like for example XSS, SQL Injection and SSRF.
However, APIF is not able to handle OpenAPI schemas, and it requires first to capture live data via a reversed proxy.

``Mass assignment'' is one of the most common security vulnerabilities in APIs, according
to OWASP'2023 report.\footref{owasp-api}
This issue happens when an attacker can update object properties that they should not have access to.
The authors of RestTestGen~\cite{viglianisi2020resttestgen} provided an extension to it to enable detection of this kind
of vulnerability~\cite{corradini2023automated}.

The work in~\cite{pasca2025llm} presents an LLM-based approach to generate tests aimed at
Broken Object Level Authorization (BOLA) vulnerabilities in REST APIs.
However, it is unclear how the automated oracles are designed.
In particular,
``\emph{this does not represent the detection of actual BOLA vulnerabilities \ldots
The potential for generating tests that incorrectly identify non-existent vulnerabilities
needs further investigation}''~\cite{pasca2025llm}.

An alternative approach to detect BOLA vulnerabilities in REST APIs has been presented in~\cite{santos2025automated}.
%There, 
In this approach, OpenAPI schemas are transformed into Petri nets, and then server logs are used to analyze possible BOLA vulnerabilities.
However, a limitation is that it
``\emph{requires having an OpenAPI specification with links. Unfortunately, this field is not widely used}''~\cite{santos2025automated}.

Our work is significantly different from this existing literature.
In our approach, we define a series of automated oracles aimed at detecting  faults related to access control.
The only requirements are the presence of an OpenAPI schema describing the API, and login information for at least two different users.
Then,  
given a fuzzer that generates a set of test cases maximizing different criteria (e.g., schema and code coverage),
we apply a post-processing strategy to generate new test cases designed to evaluate those automated oracles.
%In this paper, we define a series of automated oracles aimed at detecting  faults related to access control.
%Given a fuzzer that generates a set of test cases maximizing different criteria (e.g., schema and code coverage),
%we apply a post-processing strategy to generate new test cases based on those, aimed at checking our automated oracles.
%The only requirements are the presence of an OpenAPI schema describing the API, and login information for at least two
%different users.
With our approach, no further manual effort or custom schema annotation is required.
As far as we know, none of these automated oracles presented in this paper have been proposed and evaluated in the literature
so far.

Note that, in this work, we also discuss how we extended \evo to do SQLi and XSS attacks.
Those are not novel, as based on existing malicious payloads from the
literature,\footnote{\url{https://github.com/minimaxir/big-list-of-naughty-strings}}
and they have been evaluated already in tools such as VoAPI2~\cite{du2024vulnerability}.
However, a clear difference here is that in our approach we are able to generate \emph{executable} test cases (e.g., in JUnit format) to enable testers/developers to verify and debug any found vulnerability.

%%%%%%%%%%%%%%%%%%%%%%%%%%%%%%%%%%%%%%%%%%%%%%%%%%%%%%%%%%%%%%%%%%%%%%%%%%%%%%%%%%%%%%%%%%%%%%%%%%%%%%%%%%%%%%%%%%%%%%%
\section{Security Oracles}
\label{sec:oracles}

In this paper, we define seven new automated oracles to detect different kinds of faults related to authorization and access control, as well as two further oracles related to executing SQLi and XSS attacks.
To detect these faults, we need to generate test cases with specific scenario patterns, and then verify if the HTTP status
codes and body responses we get back from the tested API might reveal the presence of these faults.

To enable some of these security oracles,
authentication information for at least two different users are required by the fuzzer.
Better if more (at least one per different role), especially considering different access policies they might have (e.g., administrators).
No assumption on their relative access policies is needed, and no formal description of these policies is needed either.
Such information is inferred dynamically based on the 401 (not authenticated) and 403 (not authorized) HTTP status codes received
back during the fuzzing session.
The functionalities of an API can be expressed with an OpenAPI\footnote{\url{https://swagger.io/specification/}}
schema, specifying which endpoints (combination of URL \emph{paths} and HTTP \emph{verbs}) are available,
and what type of data they expect as input (e.g., query parameters and JSON body payloads).
However, OpenAPI schemas are not able to express information about access policies.

To simplify the explanations of how our novel automated oracles work, in our examples we will provide HTTP call sequences in
the following format:

\begin{lstlisting}
(AUTH) VERB PATH  -> CODES
\end{lstlisting}

Like for example a test case $T_k$ could be:

\begin{lstlisting}
(A) POST /items     -> 201
(A) GET  /items/42  -> 200
(B) GET  /items/42  -> 403
\end{lstlisting}

In parentheses we have the user (where ``?'' means it does not matter),
followed by the verb (e.g., \texttt{GET}) and the URL path, with then the resulting
HTTP status code(s) of making this call.
For simplicity, here we are omitting any query parameters, headers and body payloads.
How those are handled will be discussed next.
In this 3-step test case example $T_k$, a user $A$ creates (status 201) a new item via a \texttt{POST} request, and then
successfully accesses it with a \texttt{GET} request (status 200).
However, another user $B$ fails to access the same resource created by $A$ (status 403) as they are not authorized.

To obtain calls on specific endpoints (e.g., \texttt{POST /items}) returning some specific HTTP status codes (e.g., 201),
there might be the need for the right input data in the query parameters, body payloads and headers.
If some data is not correct (e.g., according to some defined constraints), then the API most likely would return a
4xx user error status code.
As such data and constraints could be arbitrarily complex, it is not necessarily the case that a fuzzer is able to generate such HTTP calls.
When evaluating our novel automated oracles, we need to create test scenarios with specific HTTP call sequences, where each call might
require to return some specific HTTP status code.
To achieve this, we use and copy valid test cases generated during the fuzzing session.
If for example a fuzzer is left running for 1-hour, and it generates $n$ test cases $T$ (with $|T|=n$) as output (based on maximizing some coverage criteria),
we use those $n$ test cases $T_i$ as our initial pool to construct our new security tests.

Assume we need a call on \texttt{GET /items/id} that returns a 200 status code.
This might not be trivial to obtain, if the call to generate such resource (e.g., via a \texttt{POST} or a \texttt{PUT}) requires
complex input data.
At the end of the fuzzing session, we can check if in the set $T$ there is any test case with such an HTTP call.
Assume the previous example $T_k$ was generated.
That would fit our needs.
Nevertheless, all calls \emph{after} our target call would not be necessary.
Those unnecessary calls can be \emph{sliced} away.
However, calls \emph{before} the target might or might not be needed, as they might modify the state of the API.
Hence, it cannot be determined with certainty whether they are required.
In this particular case, the first call \texttt{POST /items} creates the resource used in our target call \texttt{GET /items/id}.
We can therefore make a copy $C_k$ of $T_k$, and remove all calls after the target, but leave as they are all calls before.
In this case, $C_k$ would be:

\begin{lstlisting}
(A) POST /items     -> 201
(A) GET  /items/42  -> 200
\end{lstlisting}

When generating security tests that identify the presence of faults, it is of paramount importance that users are properly informed
about it.
\evo can output test cases in different formats, including Java, Kotlin, JavaScript and Python.
When a test case identifies a security fault, we make sure that in the generated test cases a comment is generated and
added before the HTTP call triggering/showcasing the fault.

In the following subsections, we present in detail  the nine oracles that are designed to detect the security faults.
These include
\emph{not recognized authentication} (Section~\ref{sec:F205}),
\emph{existence leakage} (Section~\ref{sec:F204}),
\emph{missed authorization checks} (Section~\ref{sec:F206}),
\emph{anonymous modifications} (Section~\ref{sec:F901}),
\emph{ignored anonymous access} (Section~\ref{sec:F900}),
\emph{leaked stack traces} (Section~\ref{sec:F902}),
and
\emph{hidden but accessible endpoints} (Section~\ref{sec:F903}),
together with input-related vulnerabilities such as
\emph{SQL injection} (Section~\ref{sec:F200}) and
\emph{cross-site scripting} (Section~\ref{sec:F201}).
These oracles form a comprehensive detection for API vulnerabilities, covering multiple facets of API security, from authentication and authorization to information disclosure and input validation.

Moreover, to better explain these novel automated oracles, we will provide some examples of artificial APIs with seeded faults, on which we show what kinds of test cases we can be automatically generated.
Those are written in Kotlin, using the popular SpringBoot framework.
All these examples are open-source and also used internally in \evo as E2E tests (and so part of its code repository), run as part of its Continuous Integration~\cite{arcuri2023building}, to verify \evo's own correctness (e.g., it should not crash, and its generated tests should be able to be compiled and run without issues).

Note that the code of the test cases that we will show are as generated by \evo version 5.1.0, \emph{as they are}.
No manual modification of any kind was done on any of these test cases for readability concerns in this paper.

%----------------------------------------------
\subsection{Not Recognized Authentication}
\label{sec:F205}

Typically, if an HTTP request is made without any authentication information, and the called endpoint requires authentication,
then the API would answer with a 401 status code.
If the user is authenticated, but, if they try to access a resource that does not belong to them, then the API should return
a 403 status code (authenticated but not authorized).

If a user makes an authenticated call, but if then they receive a 401 status code (instead of for example a 403), then it is a fault.
This is not a \emph{security} vulnerability, but rather a \emph{usability} fault.
Based on the returned 401, the user might wrongly believe that their authentication info is wrong or outdated, and waste time in
trying to resolve such issue (e.g., by contacting the IT support of the API).
As such, it would be best to fix this kind of usability fault.

In a fuzzer, simply flag as fault any 401 returned status code with an authenticated user would be unwise, as it could create
many false-positives.
For example, if a tester misconfigures the fuzzer, and provide a wrong user info, or such auth info has expired, then
false-positive faults could be identified on each single API endpoint.
Depending on the size of the API, this could result in hundreds of wrongly identified faults that could make harder to
highlight any found real fault.

To avoid this kind of false-positives, we designed the following strategy.
First, for each endpoint $X$ in the OpenAPI schema, we check if the fuzzer has generated any test case $T_1$ with an authenticated user
 (e.g., $A$) in which a 401 was returned.
If so, $X$ is potentially faulty.
We then check if authentication information for $A$ was correct.
To do so, we check if the fuzzer has generated any test $T_2$ in which $A$ was used on an endpoint $Y \neq X$ in which
a 2xx, success status code was returned (i.e., no 401 for $A$).
This is a first step to check the validity of $A$, as $Y$ could be an open endpoint in which no authentication check is
required.
To verify this, we check if in the generated tests of the fuzzer there is any test $T_3$ in which any call on $Y$
resulted in either a 401 or 403.
If so, it would mean that $Y$ requires authentication ($T_3$), the credentials of $A$ on $Y$ are correct as we get a
2xx status code ($T_2$), while the call on $X$ with $A$ identifies a fault as a wrong 401 is returned ($T_1$).

For each endpoint in the OpenAPI schema, if these three conditions are met, then we create a new
test $K=\{C_3,C_2,C_1\}$, where $C_i$ is a copy of test $T_i$, where all unnecessary calls after the target are sliced away,
as previously explained.

\begin{figure}
\begin{lstlisting}[language=Java,numbers=left,xleftmargin=2em]
/**
* Calls:
* 1 - (201) PUT:/api/resources/{id}
* 2 - (403) PUT:/api/resources/{id}
* 3 - (201) PUT:/api/resources/{id}
* 4 - (401) POST:/api/resources/
* Found 1 potential fault of type-code 205
*/
@Test @Timeout(60)
fun test_1_postOnResourcesAuthenticatedButWronglyToldNot()  {

    given().accept("*/*")
            .header("Authorization", "BAR") // BAR
            .header("x-EMextraHeader123", "")
            .put("${baseUrlOfSut}/api/resources/721")
            .then()
            .statusCode(201)
            .assertThat()
            .body(isEmptyOrNullString())

    given().accept("*/*")
            .header("Authorization", "FOO") // FOO
            .header("x-EMextraHeader123", "")
            .put("${baseUrlOfSut}/api/resources/721")
            .then()
            .statusCode(403)
            .assertThat()
            .body(isEmptyOrNullString())

    given().accept("*/*")
            .header("Authorization", "FOO") // FOO
            .header("x-EMextraHeader123", "_EM_43_XYZ_")
            .put("${baseUrlOfSut}/api/resources/577")
            .then()
            .statusCode(201)
            .assertThat()
            .body(isEmptyOrNullString())

    // Fault205. Wrongly Not Recognized as Authenticated.
    given().accept("*/*")
            .header("Authorization", "FOO") // FOO
            .header("x-EMextraHeader123", "")
            .post("${baseUrlOfSut}/api/resources/?EMextraParam123=_EM_45_XYZ_")
            .then()
            .statusCode(401)
            .assertThat()
            .body(isEmptyOrNullString())
}
\end{lstlisting}
\caption{\label{fig:801}
Example of generated test for an artificial API in which a not-recognized authentication fault is correctly identified.
}
\end{figure}

Figure~\ref{fig:801} shows an example of generated test for an artificial API  we developed to verify if our novel oracles
can identify this kind of faults.
Two users are set up: \texttt{FOO} and \texttt{BAR}.
According to the internal access policies of the API, the user \texttt{FOO} has no rights to call
\texttt{POST /api/resources/}.
However, instead of getting a 403 response, the last call in that test returns a wrong 401.
The previous call shows that the authentication info for \texttt{FOO} was correct, i.e., this detected fault is not a false-positive
due to a misconfiguration of the authentication information provided to the fuzzer.

%----------------------------------------------
\subsection{Existence Leakage}
\label{sec:F204}

Assume that the fuzzing process generated at least one test case on an endpoint definition $X$
in which a 403 status code is returned, and one test case in which a 404 is returned.
The two test cases do not need to use the same or different users to get these results.
For example, we could have $X$ being something like \texttt{/data/{id}} where a \texttt{GET} call
on \texttt{/data/42} returns a 403 status code, and a call on \texttt{/data/77} returns 404.
These two calls could be done with the same user, or with two different users.

Such scenario is problematic, as it leads to information leakage on the existence of resources.
If a user has no access to some resources (i.e., 403), they should not be able to learn which resources exist (403) and which
ones do not exist (404), as such information can be used to narrow down follow up security attacks targeted at
existing protected resources.
The HTTP standards in RFC9110\footnote{\url{https://www.rfc-editor.org/rfc/rfc9110.html}}
acknowledge the issue and explicitly states:
``\emph{An origin server that wishes to ``hide'' the current existence of a forbidden target resource MAY instead
respond with a status code of 404 (Not Found)}''.
For non-authorized users, to avoid information leakage either the APIs should always return 403 or always 404,
regardless of whether the resource exists or not.

However, there are cases in which a 404 is not leading to information leakage.
This happens when the requested resource is a subresource that belongs to the user.
Consider a new URL path $Y$, like for example \texttt{/data/{id}/bar}.
Considering  \texttt{/data/42/bar} returning 403,
having \texttt{/data/77/bar} returning 404 with an authenticated user $A$ does not necessarily mean a leakage.
It could happen for example that a \texttt{GET} on  \texttt{/data/77} returns 200, i.e., user $A$ owns it and all its
sub-resources like for example \texttt{/data/77/bar}.
But what if \texttt{/data/77} does not exist (404)?
Or the user $A$ has no access to it (403)?
Or $A$ is actually no user (i.e., making a call with no authentication information, which should had rather led to a 401)?
Then, we would have an information leakage.

To test this oracle, for each \texttt{GET} endpoint in the OpenAPI
we need to create a new test representing the following scenario:

\begin{lstlisting}
(?) GET /P/X  -> 403
(A) GET /P/X  -> 404
(A) GET /P    -> 403,404
\end{lstlisting}

Here we have two steps, with an optional third step.
First, for each \texttt{GET} endpoint we check if we have at least one test case ($T_1$) returning 403 and one ($T_2$) returning 404.
If yes, we can instantiate this scenario.
We first make a copy of $T_1$ (including all input data, like query parameters and body payloads),
and slice away all calls after that target \texttt{GET /P/X}.
A test case is composed of one or more HTTP calls.
As previously discussed, calls after our target are not needed for our oracle.
However, calls before the target might be required, as they might set the state of the API (e.g., by creating the resource).
For example, $T_1$ could be something like:

\begin{lstlisting}
(B) POST   /data  	-> 201
(C) GET    /data/42 -> 403
(B) GET    /data/42 -> 200
(B) DELETE /data/42 -> 204
\end{lstlisting}

Then, after the slice, our copy $C_1$ would be:

\begin{lstlisting}
(B) POST /data     -> 201
(C) GET  /data/42  -> 403
\end{lstlisting}

Then, we apply the same process to $T_2$: we make a copy $C_2$, where we slice all calls after the target.
For example, assume we have $T_2$ being:

\begin{lstlisting}
(A) GET  /data/77  -> 404
(A) POST /data 	   -> 201
\end{lstlisting}

Then, the copy $C_2$ would be just the first call.
Then, we can create a new test $K$ by combining $C_1$ and $C_2$:

\begin{lstlisting}
(B) POST /data     -> 201
(C) GET  /data/42  -> 403
(A) GET  /data/77  -> 404
\end{lstlisting}

One possible issue here is that the execution of $C_1$ might have side effects on the execution of $C_2$, as it might change
the state of the API (e.g., creating new data in the database, if any).
This is not a problem when executing tests in isolation, as state-of-the-art tools such as \evo can automatically reset
the state of databases after executing each test~\cite{arcuri2020sql}.
This means that we must execute $K$, and verify that, for each call,
%we get returned the same HTTP status codes as
we receive the same HTTP status codes as those returned
in the original $T_1$ and $T_2$ tests.
If the status codes remain the same, then we have detected a potential security-related fault.

However, we are not done, as we need to check if the 404 in the last call was valid, i.e., if the user owns
any ancestor resource.
If there is no authenticated user doing such call (i.e., in our example if $A$ represents no user), then there is nothing
more to check, and the fault is identified.
Otherwise, given the called target endpoint, i.e., \texttt{/data/{id}} in this case, we check in the OpenAPI the top
ancestor resource (in the URL  path) that has a \texttt{GET} endpoint.
If there is none, then there is nothing more to check, and the fault is identified.
Otherwise, we create a new \texttt{GET} call on this ancestor resource using the same user authentication (e.g., $A$ in
this example).
If we still get a 404, then the fault is identified.
If not, the result remains inconclusive.

\begin{figure}
\begin{lstlisting}[language=Java,numbers=left,xleftmargin=2em]
/**
* Calls:
* 1 - (201) PUT:/api/resources/{id}
* 2 - (403) GET:/api/resources/{id}
* 3 - (404) GET:/api/resources/{id}
* Found 1 potential fault of type-code 204
*/
@Test @Timeout(60)
fun test_5_getOnResourcAllowsUnauthorizedAccessToProtectedResource()  {

    given().accept("*/*")
            .header("Authorization", "BAR") // BAR
            .header("x-EMextraHeader123", "")
            .put("${baseUrlOfSut}/api/resources/12")
            .then()
            .statusCode(201)
            .assertThat()
            .body(isEmptyOrNullString())

    given().accept("*/*")
            .header("Authorization", "FOO") // FOO
            .header("x-EMextraHeader123", "")
            .get("${baseUrlOfSut}/api/resources/12")
            .then()
            .statusCode(403)
            .assertThat()
            .body(isEmptyOrNullString())

    // Fault204. Leakage Information Existence of Protected Resource.
    given().accept("*/*")
            .header("Authorization", "BAR") // BAR
            .header("x-EMextraHeader123", "")
            .get("${baseUrlOfSut}/api/resources/339")
            .then()
            .statusCode(404)
            .assertThat()
            .body(isEmptyOrNullString())
}
\end{lstlisting}
\caption{\label{fig:800}
Example of generated test for an artificial API in which a leakage information fault is correctly identified.
}
\end{figure}

Figure~\ref{fig:800} shows an example of generated test for an artificial API we developed to verify if our novel oracles
can identify this kind of faults.
Two users are set up: \texttt{FOO} and \texttt{BAR}.
When \texttt{FOO} accesses a resource that belongs to \texttt{BAR}, they get a 403 unauthorized response.
However, accessing a non-existing resource on the same endpoint leads to a 404, which leads to information leakage.

%----------------------------------------------
\subsection{Missed Authorization Checks}
\label{sec:F206}

Resources should be protected from unauthorized access.
For example, a user of a bank should not be able to do transactions on the accounts belonging to other users.
However, access policies could be arbitrarily complex.
Simply stating that a user should not access or manipulate resources of other users would just result in many false-positive
faults.

The most obvious example is administrator users.
A tester using a fuzzer could provide authentication information for some basic users as well as for an administrator one.
This latter would be essential for testing endpoints that only administrators can have access to.
An administrator would likely be able to access and manipulate resources of other users (albeit possibly still with some
constraints).
However, there are plenty of cases in which more complex scenarios would arise.
For example, in a web application for discussion forums, the ``moderator'' for a specific forum might not have administration
rights on the whole application.
However, they might have special rights on all the messages created by other users in that forum (e.g., to delete them if they
violate the terms of access of the application).
Another example is websites related to children (e.g., government benefits or application for kindergarten), where groups of
users (e.g., guardians for a child, typically their parents) share the same rights and resources,
but cannot access the resources of other groups (i.e., other families).
%And so on.
While these scenarios illustrate only a few examples, similarly complex access control rules can exist in various contexts.

Ideally, if there was a formal specification of these access policy rules, a fuzzer could use such specifications as automated oracle,
i.e., by checking if any rule is violated during the fuzzing session (e.g., getting a resource with a 2xx whereas it was supposed
to get a 403).
Unfortunately, we are aware of no existing formal specification for access policies in REST APIs which is widely used.
Some languages exist, like for example
XACML\footnote{\url{https://groups.oasis-open.org}}
(used for example in~\cite{martin2006defining}),
but we have never encountered any REST API using those formal specifications for access policies.
Some languages/tools such as
Open-Policy-Agent\footnote{\url{https://github.com/open-policy-agent/opa}}
and
Cerbos\footnote{\url{https://github.com/cerbos/cerbos}}
seem to have some degree of popularity, but none of the open-source APIs we found, or industries we collaborate with, make use of them.

Without a formal specification telling us what are the access policies for a REST API, how can a fuzzer identify that a resource should
had not been accessed?
For example, if a user $A$ accesses resource $X$ and they get a 200 HTTP status code, how can a fuzzer tell that being a fault (i.e., a 403
was expected with no returned data)?
This is an instance of the more general oracle problem: given $f(x)=y$, how can we be sure that $y$ is the correct output for software $f$
given the input $x$?
If $f$ crashes, then clearly we have identified a  fault for input $x$.
But what about all other cases?
The ``oracle problem'' is indeed one of the major challenges in software testing research~\cite{barr2015oracle}.

In our particular case, we use the following strategy to address the oracle problem.
In REST APIs, there are three distinct HTTP operators that can manipulate a specific resource:
\texttt{PUT} (for whole updates), \texttt{PATCH} (for partial updates) and \texttt{DELETE} (for removing the resource).
Our \emph{hypothesis} is that, if any two of these operations are forbidden (i.e., 403), but the third is allowed (i.e., 2xx),
then it is most likely a misconfiguration of the API.
For example, an authorization check on an endpoint could had been forgotten.
This would be a \emph{major} security vulnerability, with possibly catastrophic consequences.

The intuition about this oracle is rather straightforward.
Would it make any sense for a user to be forbidden to do any modification on a resource (e.g., with \texttt{PUT} and \texttt{PATCH}), but
then be allowed to just delete it from the system with a \texttt{DELETE}?
Likewise, would it make any sense that a user is not allowed to \texttt{DELETE} a resource, but then be allowed to do any kind of changes on
it that they want with a \texttt{PUT} or \texttt{PATCH}?
Or could it be just that an authorization check is wrongly missing?
Among all possible APIs and different kinds of access policy rules, there could be cases in which these scenarios could happen
and be correct.
As such, we cannot exclude the presence of some false-positives when using these rules as automated oracles.
Still, as they can be used to detect critical security faults, a non-zero amount of false-positive identified faults could be tolerable
in practical contexts.

Our approach works as following.
For each path $X$ and each one of the verb $V$ such as   \texttt{DELETE}, \texttt{PUT} and \texttt{PATCH},
we check if it exists in the OpenAPI schema (not all defined paths  might support all different HTTP verbs, such as for example \texttt{PATCH}).
If so, we check in the generated tests $T$ of the fuzzer if there is any $T_k$ for which on that verb/path $V:X$ we get a 403 response.
If $T_k$ exists, we make a copy $C_k$ in which all calls after the 403 are sliced away.
If not, we try to create it.
This is done by first checking for a test $T_c$ that creates the resource at $X$, and then add new call with a different user
executing $V:X$ on it (due to its length and complexity, we do not provide here full details of all the edge cases dealing with such resource creation;
we refer the interested reader to check our open-source implementation).
If this latter call returns a 403, then such instantiated test will be our $C_k$.
If not, our oracle is not applicable for $V:X$.

Given $C_k$ in which the last call returns a 403 with a user $A$ on path $X$ for verb $V$,
we need to add a new call $V':X$ with same user $A$ with a different verb $V'$ out of the considered three.
For example, if $V$ is \texttt{DELETE}, for $V'$ we consider  \texttt{PUT} and \texttt{PATCH}.
As making a success call with $V'$ might require specific input data (e.g., query parameters and body payloads),
we check if in $T$ there is any existing $T_j$ satisfying all these criteria.
If not, we cannot apply our oracle on this path $X$.
Otherwise, we make a copy $C_j$ of $T_j$, where all calls \emph{before} and after the target $V':X$ are sliced away, i.e.,
$C_j$ is composed of only a single call.
Then, there are two further steps.

First, in $C_j$ we modify the authentication information to rather use $A$ (in case it was using a different user).
Second, we need to make sure that both $C_k$ and $C_j$ handle the same resource.
For example, $C_k$ could deal with \texttt{/items/42} whereas $C_j$ with \texttt{/items/77}.
We create a new test $Z$ which is the concatenation of $C_k$ and $C_j$, where the resource path of $C_j$ is \emph{bound}
to be the same as in $C_k$.
For example, given $C_k$ being:

\begin{lstlisting}
(B) POST   /items     -> 201
(A) DELETE /items/42  -> 403
\end{lstlisting}

and $C_j$ being:

\begin{lstlisting}
(C) PUT /items/66 	 	-> 200
\end{lstlisting}

then $Z$ would be:

\begin{lstlisting}
(B) POST   /items     -> 201
(A) DELETE /items/42  -> 403
(A) PUT    /items/42  -> 200
\end{lstlisting}

The binding is done by either modifying the variables in the path elements of the endpoint URL, or by using the same dynamic
updates used in the tests when resources are generated dynamically.
For example, a resource could be created with a \texttt{POST} request, where the id of the newly generated resource could
be dynamic and non-deterministic.
The id could be then returned in the response of the \texttt{POST} request, e.g., in the \texttt{location} header or
as a field in the body payload.
Such id needs to be extracted, and then used in the following HTTP calls for referring to the same resource.
To achieve this, we simply rely on the existing algorithms in \evo to bind endpoints on same resources~\cite{arcuri2025tool}.

Once $Z$ is created, we execute it, and verify if the status codes are as expected, i.e., the \texttt{DELETE} returns a 403,
followed by a \texttt{PUT} or \texttt{PATCH} on same resource with same user returning a 2xx.
If this happens, then we have identified a potential security vulnerability.

\begin{figure}
\begin{lstlisting}[language=Java,numbers=left,xleftmargin=2em]
/**
* Calls:
* 1 - (201) PUT:/api/forbiddendelete/resources/{id}
* 2 - (403) DELETE:/api/forbiddendelete/resources/{id}
* 3 - (204) PUT:/api/forbiddendelete/resources/{id}
* Found 1 potential fault of type-code 206
*/
@Test @Timeout(60)
fun test_7_putOnResourcMissedAuthorizationCheck()  {

    given().accept("*/*")
            .header("Authorization", "BAR") // BAR
            .header("x-EMextraHeader123", "")
            .put("${baseUrlOfSut}/api/forbiddendelete/resources/214")
            .then()
            .statusCode(201)
            .assertThat()
            .body(isEmptyOrNullString())

    given().accept("*/*")
            .header("Authorization", "FOO") // FOO
            .header("x-EMextraHeader123", "")
            .delete("${baseUrlOfSut}/api/forbiddendelete/resources/214")
            .then()
            .statusCode(403)
            .assertThat()
            .body(isEmptyOrNullString())

    // Fault206. Allowed To Modify Resource That Likely Should Had Been Protected.
    val res_2: ValidatableResponse = given().accept("*/*")
            .header("Authorization", "FOO") // FOO
            .header("x-EMextraHeader123", "")
            .put("${baseUrlOfSut}/api/forbiddendelete/resources/214")
            .then()
            .statusCode(204)
            .assertThat()
            .body(isEmptyOrNullString())
    val location_id__32 = res_2.extract().header("location")
    assertTrue(isValidURIorEmpty(location_id__32));
}
\end{lstlisting}
\caption{\label{fig:802}
Example of generated test for an artificial API in which a missed authorization check fault is correctly identified.
}
\end{figure}

Figure~\ref{fig:802} shows an example of generated test for an artificial API we developed to verify if our novel oracles
can identify this kind of faults.
Two users are set up: \texttt{FOO} and \texttt{BAR}.
\texttt{BAR} creates a resource via a \texttt{PUT}.
A different user \texttt{FOO} is forbidden from deleting this resource (403 on \texttt{DELETE}).
However, this user can completely replace the resource with a \texttt{PUT} (as it gives a 204).

%--------------------------------------------------------------
\subsection{Anonymous Modifications}
\label{sec:F901}

The lack of authentication checks is one of most common type of security vulnerabilities.\footref{foot:42crunch}
However, it is not the case that all APIs, or all of their endpoints, must be secured.
A read-only API (e.g., enabling just \texttt{GET} requests) might provide data without need of authentication credentials (e.g., an API for a web store, where items can be browsed before needing to login to buy them).
Simply flagging as a vulnerability a \texttt{GET} endpoint, that returns a successful 2xx status code on a non-authenticated request, would just lead to false positives.

Also, it might be possible for APIs to create new data (e.g., via a \texttt{POST} call) without the need of authentication.
A common example is when registering a new user in a system, or discussion forums that allow adding anonymous comments.
Without any other available information, it is not really possible for a fuzzer to reliable decide whether \texttt{GET}/\texttt{POST} endpoints must require authentication or not.

However, this is not the case for modification operations, such as \texttt{DELETE}, \texttt{PUT} and \texttt{PATCH}.
If anonymous (i.e., non-authenticated) \texttt{DELETE} operations are allowed, then a user could delete all existing data, and delete data of other users as soon as they create new entries.
If anonymous \texttt{PUT}/\texttt{PATCH} operations are allowed, then as soon as a user creates some data, immediately another user could modify them, which would likely be detrimental for the original user.

Of course, to delete/modify the data of other users, somehow an attacker would need to determine the ids of those resources belonging to these other users.
Depending on the complexity of those ids (e.g., incremental counters or UUIDs), these ids could be predicted, or directly retrieved if there are \texttt{GET} operations on collections of resources (e.g., to \texttt{GET} all the comments in a discussion forum).

Based on these intuitions, we can flag as potential security vulnerabilities any \texttt{DELETE}, \texttt{PUT} or \texttt{PATCH} endpoint for which we get a 2xx response without any valid authenticated user information.
It might not be possible to be 100\% sure that those are indeed vulnerabilities,
i.e., for some special APIs those could be false positives.
However, considering how common the issue of missing authentication is in industrial practice,\footref{foot:42crunch} those cases are definitively something worth to flag as potential issues to manually verify/review.

In contrast to the other oracles discussed previously, to flag this type of \emph{Anonymous Modifications}, there is no need to create any special new test.
We can simply check the tests generated during the fuzzing, and see, for each  \texttt{DELETE}/\texttt{PUT}/\texttt{PATCH} endpoint, if any test was created in which a 2xx response was obtained without authentication credentials.
An edge case is \texttt{PUT} returning a 201 (meaning resource is created).
Besides doing modifications, by HTTP specs it is possible for a \texttt{PUT} to create a new resource.
But, in such cases, the API is supposed to return a 201 status code.
Therefore, we ignore those cases, i.e., anonymous \texttt{PUT} returning a 201 are not flagged as a security vulnerability.

\begin{figure}
\begin{lstlisting}[language=Java,numbers=left,xleftmargin=2em]
/**
* Calls:
* (204) PUT:/api/resources/{id}
* Found 1 potential fault of type-code 901
*/
@Test @Timeout(60)
fun test_12_putOnResourcAnonymousModifications()  {

    // Fault901. Anonymous Modifications.
    given().accept("*/*")
            .header("x-EMextraHeader123", "")
            .put("${baseUrlOfSut}/api/resources/196?EMextraParam123=42")
            .then()
            .statusCode(204)
            .assertThat()
            .body(isEmptyOrNullString())
}
\end{lstlisting}
\caption{\label{fig:901}
Example of generated test for an artificial API in which a \texttt{PUT} was successful without any authentication info.
}
\end{figure}

Figure~\ref{fig:901} shows an example of generated test for an artificial API we developed to verify if our novel oracles
can identify this kind of faults.
A \texttt{PUT} was successful without any authentication info, and without creating any new resource (status code 204).

%--------------------------------------------------------------
% \subsection{Forgotten Authentication}
\subsection{Ignore Anonymous}
\label{sec:F900}

Automatically detecting \emph{Missing Authentication} might not be possible with 100\% certainty, as some APIs might have endpoints that do not require authentication (e.g., \texttt{GET} operation on data that is public, and not restricted to some specific users).
The use of the \emph{Anonymous Modifications} oracle (recall the previous Section~\ref{sec:F901}) would cover some of the Missing Authentication cases, but not all.

A complementary approach is to check cases in which an API does verify users' credentials, but only if those are provided.
For example, assume that on an endpoint we get a 401 or 403 when making an HTTP call with some credentials (where 401 would mean the credentials are invalid, e.g., like an expired JWT token).
If it is possible to get a 2xx on the same endpoint without any credential, then that is a clear security vulnerability.
This could happen if the API is wrongly configured to verify the validity of credentials when provided, but then doing no authentication control if no credential is provided, i.e., the validity of anonymous requests is ignored.

To generate a test case to verify such scenario for each endpoint, we do the following.
First, we check if any test case $T_1$ was created with a 401 or 403 on such endpoint for any given user $A$.
If so, we make a copy $C_1$, where all calls after the 401/403 are sliced away.
Then, we check if there was any test case $T_2$ on the same endpoint in which a 2xx was returned with no authenticated user.

If $T_2$ exists, then we have found an \emph{Ignore Anonymous} fault, which can be showcased with a new test with concatenated $C_1$ and $C_2$, where $C_2$ is a sliced copy of $T_2$.

If $T_2$ does not exist, we verify if any $T_3$ test case was generated in which the endpoint was covered with 2xx for a given different user $B$.
If so, we make a sliced copy $C_3$, where the credentials of $B$ are \emph{removed}.
If $C_3$ still lead to get a 2xx, the fault is found, which can be shown by concatenating $C_1$ and $C_3$.
For example, given $T_1$ as:

\begin{lstlisting}
(?) POST   /items     -> 201
(A) DELETE /items/42  -> 403
...
\end{lstlisting}

and $T_3$ as:

\begin{lstlisting}
(B) POST   /items     -> 201
(B) DELETE /items/66  -> 200
...
\end{lstlisting}

then, the resulting $C_1$-$C_3$ would become:

\begin{lstlisting}
(?) POST   /items     -> 201
(A) DELETE /items/42  -> 403
(B) POST   /items     -> 201
(-) DELETE /items/66  -> 200
\end{lstlisting}

Note that, simply removing the authentication credential $A$ in $T_1$ to create $C_3$ (instead of searching for a different test $T_3$ that returned 2xx) might not work.
This is because the call might require specific data (e.g., in query parameters and body payloads) to avoid 400 user error responses.
So that is the reason why we try to start from a test $T_3$ that has no input validation errors (i.e., returning 2xx).

\begin{figure}
\begin{lstlisting}[language=Java,numbers=left,xleftmargin=2em]
/**
* Calls:
* 1 - (201) PUT:/api/resources/{id}
* 2 - (200) GET:/api/resources/{id}
* 3 - (403) GET:/api/resources/{id}
* 4 - (200) GET:/api/resources/{id}
* Found 1 potential fault of type-code 900
*/
@Test @Timeout(60)
fun test_5_getOnResourcIgnoreAnonymous()  {

    given().accept("*/*")
            .header("Authorization", "FOO") // FOO
            .header("x-EMextraHeader123", "42")
            .put("${baseUrlOfSut}/api/resources/483")
            .then()
            .statusCode(201)
            .assertThat()
            .body(isEmptyOrNullString())

    given().accept("*/*")
            .header("Authorization", "FOO") // FOO
            .header("x-EMextraHeader123", "42")
            .get("${baseUrlOfSut}/api/resources/483")
            .then()
            .statusCode(200)
            .assertThat()
            .contentType("text/plain")
            .body(containsString("FOO"))

    given().accept("*/*")
            .header("Authorization", "BAR") // BAR
            .header("x-EMextraHeader123", "42")
            .get("${baseUrlOfSut}/api/resources/483")
            .then()
            .statusCode(403)
            .assertThat()
            .body(isEmptyOrNullString())

    // Fault900. A Protected Resource Is Accessible Without Providing Any Authentication.
    given().accept("*/*")
            .header("x-EMextraHeader123", "42")
            .get("${baseUrlOfSut}/api/resources/483")
            .then()
            .statusCode(200)
            .assertThat()
            .contentType("text/plain")
            .body(containsString("FOO"))
}
\end{lstlisting}
\caption{\label{fig:900}
Example of generated test for an artificial API in which a protected resource is accessible without providing authentication.
}
\end{figure}

Figure~\ref{fig:900} shows an example of generated test for an artificial API we developed to verify if our novel oracles
can identify this kind of faults.
Two users are set up: \texttt{FOO} and \texttt{BAR}.
\texttt{FOO} creates a resource via a \texttt{PUT}, and it is able to successfully retrieve it with a \texttt{GET}.
A different user \texttt{BAR} is forbidden from accessing this resource (403 on \texttt{GET}).
However, trying to access such protected resource without any authentication info is successful (i.e., 200 on \texttt{GET}).

%--------------------------------------------------------------
\subsection{Leaked Stack Traces}
\label{sec:F902}

Software faults are common.
In a REST API, when an application crash due to a fault (e.g., a null-pointer exception), then typically only the application layer of the API fails, and a 500 HTTP status code is sent back to the client (i.e., the entire HTTP server does not crash, so the TCP connection is still open and an error message can be sent back).

To help debugging the fault, the server could log the \emph{stack trace} of the exception to a log file, and/or send it back in the body payload of the HTTP 500 response.
Where this latter behavior can be useful during the development of an API (e.g., while an engineering is developing a new feature, like a new endpoint), it can be a major security vulnerability for APIs run in production.
An attacker that triggers a 500 fault (e.g., by using a fuzzer) would be able to see such stack traces.
From the stack traces they could infer the programming language and framework used to develop the API.
Such information could then be used for more targeted attacks (e.g., if that specific framework has known zero-day vulnerabilities).

As a rule of thumb, HTTP 500 status responses should provide no information to the clients about any internal details of the API, as such information can be exploited by malicious actors.
Developers could enable stack traces in the responses during development, and, then, forget to disable such feature when the API is deployed to production.
This is a security vulnerability.

To verify whether a HTTP 500 response contains stack traces, we developed a series of regular expressions targeted at all the major programming languages (e.g., Java, Python, C\#, JavaScript, Go, PHP, Ruby and Rust).
Stack traces could be returned in plain-text, or embedded in a JSON response.
Both cases are handled.

As such text analyses have a non-negligible computational cost, they are not executed on each HTTP 500 response.
During a 1-24 hour fuzzing campaign, hundreds of thousands of such tests could be evaluated.
Once the fuzzing session is completed, for each endpoint we check if any test was generated returning a 500.
If so, the text analysis is applied only on that test for that endpoint.
Given $N$ different endpoints in the API, these text analyses are run at most $N$ times on $N$ different tests.

When doing fuzzing with \evo, by default the \emph{Leaked Stack Traces} is activated.
However, in case of its use by developers that willingly want to have stack traces in their HTTP responses, such oracle should be deactivated, as it would otherwise generate only false positives.
\evo is a highly tunable fuzzer, with hundreds of available
(and documented\footnote{\url{https://github.com/WebFuzzing/EvoMaster/blob/master/docs/options.md}})
options~\cite{arcuri2023building}.
The use of each single oracle can be toggled via configuration settings (command-line input arguments and/or configuration files).

\begin{figure}
\begin{lstlisting}[language=Java,numbers=left,xleftmargin=2em]
/**
* Calls:
* (500) GET:/api/resources/null-pointer-json
* Found 2 potential faults. Type-codes: 100, 902
*/
@Test @Timeout(60)
fun test_1_getOnNull_pointer_jsonShowsFaults_100_902()  {

    // Fault100. HTTP Status 500. com/foo/rest/examples/spring/openapi/v3/security/stacktrace/json/StackTraceJSONApplication_34_nullPointerJson
    // Fault902. Leaked Stack Trace.
    given().accept("application/json")
            .header("x-EMextraHeader123", "")
            .get("${baseUrlOfSut}/api/resources/null-pointer-json?EMextraParam123=_EM_0_XYZ_")
            .then()
            .statusCode(500)// com/foo/rest/examples/spring/openapi/v3/security/stacktrace/json/StackTraceJSONApplication_34_nullPointerJson
            .assertThat()
            .contentType("application/json")
            .body("'error'.'type'", containsString("java.lang.NullPointerException"))
            .body("'error'.'message'", containsString(""))
            .body("'error'.'stack'.size()", equalTo(56))
            .body("'error'.'stack'", hasItems("java.lang.NullPointerException", "\tat com.foo.rest.examples.spring.openapi.v3.security.stacktrace.json.StackTraceJSONApplication.nullPointerJson(StackTraceJSONApplication.kt:36)", "\tat jdk.internal.reflect.GeneratedMethodAccessor90.invoke(Unknown Source)", "\tat java.base/jdk.internal.reflect.DelegatingMethodAccessorImpl.invoke(DelegatingMethodAccessorImpl.java:43)", "\tat java.base/java.lang.reflect.Method.invoke(Method.java:569)", "\tat org.springframework.web.method.support.InvocableHandlerMethod.doInvoke(InvocableHandlerMethod.java:197)", "\tat org.springframework.web.method.support.InvocableHandlerMethod.invokeForRequest(InvocableHandlerMethod.java:141)",
            ...
\end{lstlisting}
\caption{\label{fig:902}
Excerpt of a generated test for an artificial API in which an exception stack-trace is present in the response body.
Due to long stack-trace (56 lines), the test is not displayed in its entirety.
}
\end{figure}

Figure~\ref{fig:902} shows an example of a test case that leads to an internal crash (HTTP status code 500), in which the stack-trace ends up in the response body.
Here, two distinct faults are found and flagged: the fact that the API crashed (status code 500), and that a stack-trace leaked in the response.

%--------------------------------------------------------------
\subsection{Hidden Accessible}
\label{sec:F903}

An OpenAPI schema defines what the API can provide.
What if the API has working endpoints that are not declared in the schema?
This could be a bug in the schema itself, i.e., the designers/developers of the API forgot to specify the endpoint.
This can be the case when schemas are written manually (and not automatically derived from the source code of the API, e.g., using libraries such as SpringDoc for SpringBoot applications).
Although this is not a security issue, it is still a \emph{usability} fault that requires fixing.

However, it might also be that those endpoints are not supposed to be accessible.
They could be used just for debugging/testing, or being work-in-progress.
Having those endpoints accessible in production by users could become a security risk.
If an endpoint is ``hidden'' (i.e., not declared in the schema), but it is still ``accessible'' when clients call it directly,
we have a problem (either a usability or a security one).

If a client makes a call to an endpoint for a resource that exists, but the ``verb'' used is not supported for it, then the server should return either a HTTP 405 status code (\emph{Method Not Allowed}) or a 501 (\emph{Not Implemented}).
To limit disclosing information for security reasons, the server might even decide to return a 403 in those cases instead.
If a call is made to an endpoint that is not declared in the schema, but the response contains a status different from 403, 405 and 501, then there is a problem.

To check these properties, we use the following approach.
For each distinct resource path declared in the schema, we do make a \texttt{OPTIONS} HTTP call.
This will provide information about the available verbs for that path, returned in the \texttt{Allow} HTTP header.
We then check if there is any allowed verb not declared in the schema (excluding \texttt{OPTIONS} itself and \texttt{HEAD}).
If so, for each of those hidden verbs, we make a call with it on that resource path, with no body payloads and no query parameters (as there is no info about them in the schema).
If we obtain a result different than a 403, 405 or 501, we flag it as an error.
To reproduce this problem, we instantiate a new test case with the \texttt{OPTIONS} call, followed by the call with the hidden verb.

Technically, the call with \texttt{OPTIONS} might be redundant.
One could simply directly call all endpoints with verbs not declared in the schema, and see what is returned as status code.
The use of \texttt{OPTIONS} just makes the tests more clear, better highlighting the severity of the issue.

\begin{figure}
\begin{lstlisting}[language=Java,numbers=left,xleftmargin=2em]
/**
* Calls:
* 1 - (200) OPTIONS:/api/resources
* 2 - (200) GET:/api/resources
* Found 3 potential faults. Type-codes: 101, 903
*/
@Test @Timeout(60)
fun test_1_getOnResourcesShowsFaults_101_903()  {

    // Fault101. Received A Response From API With A Structure/Data That Is Not Matching Its Schema. Type: validation.request.operation.notAllowed OPTIONS operation not allowed on path '/api/resources'.
    given().accept("*/*")
            .options("${baseUrlOfSut}/api/resources")
            .then()
            .statusCode(200)
            .assertThat()
            // .header("Allow", "HEAD,POST,GET,OPTIONS")
            .body(isEmptyOrNullString())

    // Fault101. Received A Response From API With A Structure/Data That Is Not Matching Its Schema. Type: validation.request.operation.notAllowed GET operation not allowed on path '/api/resources'.
    // Fault903. Hidden Accessible Endpoint.
    given().accept("*/*")
            .get("${baseUrlOfSut}/api/resources")
            .then()
            .statusCode(200)
            .assertThat()
            .contentType("text/plain")
            .body(containsString("OK"))
}
\end{lstlisting}
\caption{\label{fig:903}
Example of generated test for an artificial API in which first a call is made via \texttt{OPTIONS} to determine which methods are allowed, followed by a successful call to a \texttt{GET} that is not defined in the schema.
}
\end{figure}

Figure~\ref{fig:903} shows an example of test in which first an \texttt{OPTIONS} call is made on a path resource.
Based on the \texttt{Allow} header in the response, considering that no \texttt{GET} is declared in the schema (not shown), a follow up call with a \texttt{GET} is done.
As this is successful, such call is marked as a potential security vulnerability (besides being as well a violated schema fault).

%--------------------------------------------------------------
\subsection{SQL Injection}
\label{sec:F200}

In 2025, \emph{Injection} is still one of the most common security vulnerability in web applications,\footref{owasp-web} although its presence in APIs is less prevalent.\footref{owasp-api}
In the research literature, there is a large body of knowledge regarding how to automatically detect injection vulnerabilities such as SQL Injection (e.g., \cite{KGJE09,appelt2014automated,aziz2016search,liu2020deepsqli}), including work on REST APIs~\cite{du2024vulnerability}.

In our work, we do not only try to detect security vulnerabilities, but we also aim at being able to generate executable test cases to enable engineers to effectively debug such faults.
In our approach, for each endpoint, we check if any test case $T_i$ was generated in which a successful 2xx response was obtained.
Responses that end up with user errors (i.e., 4xx) or server errors (5xx) might be less likely to involve database interactions.
Then, for each input variable that is a string (e.g., in query parameters and fields in body payloads), we append a malicious SQL injection payload.
This is as long as such extended string is still valid for any constraints expressed in the OpenAPI schema (e.g., length, enum and pattern constraints, if any is defined), as otherwise the API will just return a 400 status code without executing any business logic.

We need to make sure that we can automatically detect if a SQL Injection attack is successful, also when dealing with black-box testing where no direct access to the database is possible.
For this case, we focus on \emph{sleep} payloads, like for example
\texttt{' OR SLEEP(5)-- -}.
Given $N$ endpoints and $P$ different malicious payloads to evaluate (with $P=16$ in our case), at most $N\times P$ new test cases need to be evaluated.
In the case of white-box testing, in which we can detect if the API is using any SQL database, if no database is used, then the evaluation of this \emph{SQL Injection} oracle is automatically skipped.

If the injection of a sleep of $S$ seconds is successful, then the test case should take at least $S$ seconds longer to run.
However, execution time of a test case for a REST API is non-deterministic, as it depends on many factors such as network load, and OS CPU schedule.
Even when re-executing the same test twice, a certain amount of time variability is expected.
The injected sleep time should be long enough to minimize risks of this source of variability, but not too long to make the testing unpractical.
Ideally, HTTP calls to an API should just take few tens or hundreds of milliseconds.
In those cases, a sleep of few seconds would unlikely to be affected by noise.
But, if for any reason the execution time of HTTP calls for some endpoints are expected to be taking seconds, then such time variability might have negative effects.
To handle this, we exclude from our analysis any HTTP call that is taking longer than $M$ seconds (e.g., $M=2$).

To create test cases to find and showcase SQL Injection vulnerabilities, we do as follow.
For each endpoint, we check if any test case $T_1$ was created where an HTTP call returned a 2xx and lasted less than $M$ seconds.
If any is found, we make a copy $C_1$ in which any call after the target (let's call it $X$) is sliced away.
Then, we make a copy of the target HTTP call (let's call this copy $Y$), in which we inject malicious payloads, and append it to $C_1$.
If the target HTTP call $X$ required the ids of dynamically generated resources in the test, we need to make sure the appended copied call $Y$ refers to the same dynamic ids.
For example, given $T_1$ for the endpoint \texttt{GET /items/\{id\}}:

\begin{lstlisting}
(?) POST /items     -> 201
(?) GET  /items/42  -> 200
...
\end{lstlisting}

then the resulting $C_1$ would be:

\begin{lstlisting}
(?) POST /items     -> 201
(?) GET  /items/42  -> 200
(?) GET  /items/42  -> 200
\end{lstlisting}

When evaluating $C_1$, if $X$ takes less than $M=2$ seconds, and $Y$ takes more than $S=5$ seconds, then we have detected and can reproduced the SQL Injection vulnerability.
Note, to be able to properly convey this information to the final users (e.g., developers and testers), our output test suites are enhanced with time tracking of the HTTP execution times, with assertions based on the values $M$ and $S$.

\begin{figure}
\begin{lstlisting}[language=Java,numbers=left,xleftmargin=2em]
/**
* Calls:
* 1 - (200) POST:/api/sqli/body/vulnerable
* 2 - (200) POST:/api/sqli/body/vulnerable
* Found 1 potential fault of type-code 200
*/
@Test @Timeout(60)
fun test_3_postOnVulnerableVulnerableToSQLInjection()  {


    // res_0_ms stores the start time in milliseconds
    val res_0_ms = System.currentTimeMillis()

    given().accept("*/*")
            .header("x-EMextraHeader123", "")
            .contentType("application/json")
            .body(" { " +
                " \"password\": \"_EM_110_XYZ_\", " +
                " \"username\": \"_EM_111_XYZ_\" " +
                " } ")
            .post("${baseUrlOfSut}/api/sqli/body/vulnerable?EMextraParam123=_EM_112_XYZ_")
            .then()
            .statusCode(200)
            .assertThat()
            .contentType("text/plain")
            .body(containsString("MATCHED: 0"))

    // res_2_ms stores the total execution time in milliseconds
    val res_2_ms = System.currentTimeMillis() - res_0_ms

    // Note: No SQL Injection vulnerability detected in this call. Expected response time (sqliBaselineMaxResponseTimeMs) should be less than 2000 ms.
    assertTrue(res_2_ms < 2000)


    // res_3_ms stores the start time in milliseconds
    val res_3_ms = System.currentTimeMillis()

    // Fault200. SQL Injection (SQLi).
    given().accept("*/*")
            .header("x-EMextraHeader123", "")
            .contentType("application/json")
            .body(" { " +
                " \"password\": \"_EM_2_XYZ_\u0027 OR SLEEP(5.00)-- -\", " +
                " \"username\": \"_EM_3_XYZ_\u0027 OR SLEEP(5.00)-- -\" " +
                " } ")
            .post("${baseUrlOfSut}/api/sqli/body/vulnerable?EMextraParam123=_EM_4_XYZ_%27+OR+SLEEP%285.00%29--+-")
            .then()
            .statusCode(200)
            .assertThat()
            .contentType("text/plain")
            .body(containsString("MATCHED: 0"))

    // res_5_ms stores the total execution time in milliseconds
    val res_5_ms = System.currentTimeMillis() - res_3_ms

    // Note: SQL Injection vulnerability detected in this call. Expected response time (sqliInjectedSleepDurationMs) should be greater than 5000 ms.
    assertTrue(res_5_ms > 5000)
}
\end{lstlisting}
\caption{\label{fig:200}
Example of generated test for an artificial API in which a SQL Injection vulnerability is detected.
}
\end{figure}

Figure~\ref{fig:200} shows an example of a test case with two HTTP calls.
The first call is executed, and we make sure to verify it takes less than 2 seconds.
The second call is a copy of the first, but with some SQL Injection payloads.
Here, we verify the call takes more than 5 seconds.

%--------------------------------------------------------------
\subsection{Cross Site Scripting (XSS)}
\label{sec:F201}

Cross-Site Scripting (XSS) is a widespread type of security vulnerability where attackers could inject malicious client-side scripts (usually JavaScript) into the frontend of web applications.
As it is a common vulnerability in web applications, a significant body of research has been dedicated to analyze and design techniques to identify it and prevent it~\cite{hydara2015current}.
For example, a payload like \texttt{<img src=x onerror=alert('XSS')>} would execute an \texttt{alert} command if displayed directly in the HTML of a web page without escaping first the opening/closing tag characters.

In the case of REST APIs, there is no web frontend, though.
As such, the API itself is no subject to XSS vulnerabilities.
However, the API ``might'' store malicious XSS payloads that, if later retrieved by a faulty web frontend application, they ``might'' result in XSS exploits if not escaped.
As such, \emph{Stored XSS} payloads are a potential threat, if malicious payloads are saved as they are without any input sanitization.
This could happen with operations like \texttt{POST}, \texttt{PUT} and \texttt{PATCH}.

Besides storing malicious XSS payloads, \emph{Reflected XSS} are problematic as well.
A \texttt{GET} request with a malicious payload would be dangerous if the payload is returned as part of the response, and if the response is displayed without sanitization.
A typical example of this problem is in search web forms where the searched query is displayed with its results.
Triggering a malicious search could be as easy as clicking on a compromised link, if the searched query is passed by URL query parameters.

To create test cases to flag potential XSS vulnerabilities, we do as follow, using a similar approach as done for \emph{SQL Injection} handling.
For each endpoint, we check if any test case $T_1$ was created returning a 2xx.
We make a copy $C_1$ where we slice away any call after the target endpoint.
Given $P$ different XSS payloads, we apply them to each string input, i.e., we replace any string value (e.g., in query parameters and body payloads), but only as long as the OpenAPI constraints are satisfied (e.g., on string length, enumeration and pattern constraints).
If the execution of this new test $C_1$ still returns an HTTP 2xx, and the malicious payload is present in the body payload of this response as it is (i.e., with no sanitization), then we have detected a potential \emph{Reflected XSS}.

For the case of \emph{Stored XSS}, this approach might not be enough, as the results of \texttt{POST}, \texttt{PUT} and \texttt{PATCH} might, or might not, return any data in the body payload of their responses (e.g., a \texttt{POST} could return a representation of the created resource, including a dynamically generated id on the server-side, but that is not a requirement).
Furthermore, even if data is returned, it does not mean that such data was actually \emph{stored} in the backend system of the API.
To handle these cases, we need to append to $C_1$ a \texttt{GET} request on the same resource manipulated/created by the \texttt{POST}, \texttt{PUT} and \texttt{PATCH}.
In this \texttt{GET} call, we apply the same $P$ payloads, and verify if its execution lead to see back the any of this payload as they are, without any input sanitization.

Given $N$ endpoints, at most $N\times P$ new test cases are generated and evaluated to verify the presence of potential XSS vulnerabilities.

\begin{figure}
\begin{lstlisting}[language=Java,numbers=left,xleftmargin=2em]
/**
* Calls:
* 1 - (200) POST:/api/stored/json/guestbook
* 2 - (200) GET:/api/stored/json/guestbook
* Found 1 potential fault of type-code 201
*/
@Test @Timeout(60)
fun test_3_getOnGuestbookVulnerableToXSS()  {

    given().accept("application/json")
            .header("x-EMextraHeader123", "<img src=x onerror=alert('XSS')>")
            .post("${baseUrlOfSut}/api/stored/json/guestbook?" +
                "name=%3Cimg+src%3Dx+onerror%3Dalert%28%27XSS%27%29%3E&" +
                "entry=%3Cimg+src%3Dx+onerror%3Dalert%28%27XSS%27%29%3E&" +
                "EMextraParam123=%3Cimg+src%3Dx+onerror%3Dalert%28%27XSS%27%29%3E")
            .then()
            .statusCode(200)
            .assertThat()
            .contentType("application/json")
            .body("'message'", containsString("Guestbook Entry Stored! Thank you for signing our guestbook!"))
            .body("'success'", equalTo(true))

    // Fault201. Cross-Site Scripting (XSS).
    given().accept("application/json")
            .header("x-EMextraHeader123", "<img src=x onerror=alert('XSS')>")
            .get("${baseUrlOfSut}/api/stored/json/guestbook?EMextraParam123=%3Cimg+src%3Dx+onerror%3Dalert%28%27XSS%27%29%3E")
            .then()
            .statusCode(200)
            .assertThat()
            .contentType("application/json")
            .body("'entries'.size()", equalTo(1))
            .body("'entries'[0].'name'", containsString("<img src=x onerror=alert('XSS')>"))
            .body("'entries'[0].'entry'", containsString("<img src=x onerror=alert('XSS')>"))
}
\end{lstlisting}
\caption{\label{fig:201}
Example of generated test for an artificial API in which a stored XSS potential vulnerability is detected.
}
\end{figure}

Figure~\ref{fig:201} shows an example of generated test case in which a resource is created with a \texttt{POST}, where some XSS payloads are passed as query parameters.
This call is followed by a \texttt{GET} on the same resource, in which we verify that the XSS payload appears in the response's body.

%%%%%%%%%%%%%%%%%%%%%%%%%%%%%%%%%%%%%%%%%%%%%%%%%%%%%%%%%%%%%%%%%%%%%%%%%%%%%%%%%%%%%%%%%%%%%%%%%%%%%%%%%%%%%%%%%%%%%%%
\section{Fuzzer Integration}
\label{sec:integration}

To use our novel oracles, we need to create specific scenarios, and then verify if any fault is detected based on the
HTTP status codes and body responses we receive back from the tested API.
To check and verify authentication and authorization properties, we need to be able to generate test cases that can exercise
all the different functionalities of the API.
Input data can have complex constraints (e.g., in query parameters and body payloads),
and so test cases that trivially return 400 user errors would not be helpful for our goals.

The integration of our approach into existing fuzzers is as follows.
We apply our security validation as a \emph{post-processing} phase.
We let a fuzzer run for the amount of time specified by the users.
In academic experiments, this is typically 1 hour~\cite{golmohammadi2023testing}.
Within 1-hour, a fuzzer could evaluate hundreds of thousands of HTTP calls.
However, most likely no human test engineer would be interested in analyzing and using test suites with hundreds of
thousands of test cases~\cite{icst2025vw}.
It is simply not feasible.
As such, fuzzers typically provide as output a minimized test suite, aimed at maximizing different criteria.
These could be about covering different black-box aspects of the API~\cite{martin2019test} (e.g., based on the OpenAPI schema, like covering
different HTTP status codes for each defined endpoint), or code coverage metrics in white-box testing~\cite{arcuri2019restful}.

After the fuzzing session is over, the fuzzer would have generated a set $T$ of test cases.
These tests are then used as a base to create new test scenarios, as previously described in Section~\ref{sec:oracles}.
As such, how these $T$ tests are created is not so important, and any state-of-the-art fuzzer could be used, as long as
it can generate tests using authentication information.
Furthermore, technically there would not even be the need of a fuzzer, as existing manually written tests (if any) could be used as starting point $T$.
The technical challenge there would then be how to effectively \emph{carve} those tests to enable their manipulation and modifications for generating new test cases.

For our experiments, we used \evo, as it fits those constraints.
However, one limitation is that it was not good at generating tests in which 403 unauthorized accesses were detected.
Given a set of authentication information as input to the fuzzer (besides a copy of the OpenAPI schema for the tested API), \evo can generate
tests with and without authentication information.
However, calls with different users on the same resources (which could lead to 403 errors) were not handled.
As such, we needed to make sure to try to create such kind of tests before applying our novel oracles, as some of those rely on
existing tests with 403 responses to create the needed test scenarios.

For each verb/path $V:X$ in the OpenAPI schema,
we check if we have any test case generated in $T$ in which such a call returns a 403 (unauthorized).
If not, we try to create a new test for it and add it to $T$.
If there is no call in $T$ with a 401 (not authenticated), then most likely there is no authentication mechanism on that endpoint,
and it can be skipped.
Otherwise, we then check if there is any test $T_a$ for which a successful 2xx response is obtained with any authenticated user $A$.
If so, we make a copy $C_a$ in which all calls after the target are sliced away.
Then, for each other defined user $B$, we create a copy $C_b$ of $C_a$, and we duplicate the last call in $C_b$ by replacing $A$ with $B$.
For example, if we have $C_a$ being:

\begin{lstlisting}
(A) POST /data     -> 201
(A) GET  /data/42  -> 200
\end{lstlisting}

then $C_b$ would be:

\begin{lstlisting}
(A) POST /data     -> 201
(A) GET  /data/42  -> 200
(B) GET  /data/42  -> 403
\end{lstlisting}

If
%when we execute $C_b$ the last call returns a 403
executing $C_b$ returns a 403 from the last call, then we are done for $V:X$, and $C_b$ can be added to $T$.
Otherwise, we repeat this process for the other users (if there is more than two) and for any other tests $T_a$
that meet those criteria.

The number of new test cases we generate to evaluate security properties in our post-processing phase is linearly
dependent on the number of endpoints $N$ in the tested API, i.e., its complexity is $O(N)$.
Typically, these are at most a few hundreds.
As such, creating and evaluating these security tests takes just a few seconds or minutes, depending on the tested API.
In the context of fuzzing an API for hours (e.g., between 1 and 24 hours), in most cases the overhead of our novel techniques is practically negligible, as we will show in our empirical study in Section~\ref{sec:rq4}.

%%%%%%%%%%%%%%%%%%%%%%%%%%%%%%%%%%%%%%%%%%%%%%%%%%%%%%%%%%%%%%%%%%%%%%%%%%%%%%%%%%%%%%%%%%%%%%%%%%%%%%%%%%%%%%%%%%%%%%%
\section{Empirical Study}
\label{sec:empirical}

In this paper, we carried out an empirical study aimed at answering the following research questions:

\begin{description}
    \item[{\bf RQ1}:] Are our novel techniques able to identify injected faults in artificial examples?
    \item[{\bf RQ2}:] Are our novel techniques able to identify faults in example APIs used in the security literature?
	\item[{\bf RQ3}:] What security faults can be found in existing real-world APIs?
	\item[{\bf RQ4}:] What is the overhead cost of running our novel oracles?
\end{description}

%--------------------------------------------
\subsection{API Selection}

To answer {\bf RQ1}, we created nine small APIs, using Kotlin and SpringBoot.
In each one, we injected a security fault for each of our seven novel oracles and the two injection cases.
These APIs are simple, with no complex input constraints, or code execution flow depending on database state.
Faults might not be detected if the code in which they reside is never executed due to the complexity of generating
the right data.
Our goal here in {\bf RQ1} is to study the feasibility of our novel techniques without confounding factors.

For reason of space, we do not provide here full code of these nine example APIs.
They are provided open-source with our extension to \evo.
The examples shown in Section~\ref{sec:oracles} were based on these artificial APIs,
like the generated tests shown in 
%Figure~\ref{fig:801}, Figure~\ref{fig:800} and Figure~\ref{fig:802}.
Figures~\ref{fig:801}~--~\ref{fig:201}.

\begin{table}[!t]
\centering
\caption{ Statistics of the employed REST APIs in our empirical study for black-box testing,
including number of source files, numbers of lines of code (LOCs), the number of HTTP endpoints,
and their identifier on GitHub.
\label{tab:suts-bb}
}
\vspace{-1.5\baselineskip}
\resizebox{1.0\textwidth}{!}{
\begin{tabular}{l rrr l}\\
\toprule
SUT & \#SourceFiles & \#LOCs & \#Endpoints  & GitHub \\
\midrule
\emph{capital} &  103  &  4481 &  21  & \emph{Checkmarx/capital}\\
\emph{crapi} &  564 &  129026 &  44 & \emph{OWASP/crAPI}\\
\emph{damn-vulnerable-rest} &  75 & 4470  &  21 & \emph{theowni/Damn-Vulnerable-RESTaurant-API-Game}\\
\emph{dvapi} &  13 & 1211  &  16 & \emph{payatu/DVAPI}\\
\emph{dvws-node} &  27 & 3083  &  31 & \emph{snoopysecurity/dvws-node}\\
\emph{vampi} &  11 & 520  &  14 & \emph{erev0s/VAmPI}\\
\emph{vulnerable-rest-api} &  30 & 837  &  19 & \emph{bnematzadeh/vulnerable-rest-api}\\
\emph{webgoat} &  401 & 25972  &  201 & \emph{WebGoat/WebGoat}\\
\midrule
Total 8 & 1224 &  169600 & 367 & \\
\bottomrule
\end{tabular}
}
\end{table}

To answer {\bf RQ2}, we looked at the security research literature for existing APIs used for demonstrating different
types of security vulnerabilities.
After a search in popular resources such as OWASP, we selected eight widely used open-source APIs.
% \emph{capital},\footnote{https://github.com/Checkmarx/capital}
% \emph{crapi}\footnote{https://github.com/OWASP/crAPI}
% and
% \emph{webgoat}.\footnote{https://github.com/WebGoat/WebGoat}
Table~\ref{tab:suts-bb} shows some statistics on these eight APIs, including links to their repository on GitHub.
All these APIs are run through Docker Compose, where all of their dependencies are started as well (e.g., databases).
Note that \emph{crapi} is a microservice running different services.
All are run in our experiments.
We fuzz the whole application through its API Gateway.

Note that these APIs are deliberately vulnerable, with several different types of injected security faults.
All these APIs require authentication information (e.g., valid usernames/passwords for some existing users).
When the applications start, such info is set in their databases.
Authentication information for the fuzzer is provided in YAML files, using the Web Fuzzing Commons (WFC) format~\cite{sahin_2025_wfc}.

\begin{table}[!t]
\centering
\caption{ Statistics of the employed 36 REST APIs in our empirical study for white-box testing,
including number of source files, numbers of lines of code (LOCs), the number of HTTP endpoints,
whether they require authentication, and the databases they use (if any).
\label{tab:suts-wb}
}
\vspace{-1.5\baselineskip}
\resizebox{1.0\textwidth}{!}{
\begin{tabular}{l rrrrr}\\
\toprule
SUT & \#SourceFiles & \#LOCs & \#Endpoints & Auth& Databases\\
\midrule
\emph{bibliothek} &  33 &  2176 &  8 &  &  MongoDB \\
\emph{blogapi} &  89 &  4787 &  52 & \checkmark &  MySQL \\
\emph{catwatch} &  106 &  9636 &  14 &  &  H2 \\
\emph{cwa-verification} &  47 &  3955 &  5 &  &  H2 \\
\emph{erc20-rest-service} &  7 &  1378 &  13 &  &   \\
\emph{familie-ba-sak} &  1089 &  143556 &  183 & \checkmark &  PostgreSQL \\
\emph{features-service} &  39 &  2275 &  18 &  &  H2 \\
\emph{genome-nexus} &  405 &  30004 &  23 &  &  MongoDB \\
\emph{gestaohospital} &  33 &  3506 &  20 &  &  MongoDB \\
\emph{http-patch-spring} &  30 &  1450 &  6 &  &   \\
\emph{languagetool} &  1385 &  174781 &  2 &  &   \\
\emph{market} &  124 &  9861 &  13 & \checkmark &  H2 \\
\emph{microcks} &  471 &  66186 &  88 & \checkmark &  MongoDB \\
\emph{ocvn} &  526 &  45521 &  258 & \checkmark &  H2;MongoDB \\
\emph{ohsome-api} &  87 &  14166 &  134 &  &  OSHDB \\
\emph{pay-publicapi} &  377 &  34576 &  10 & \checkmark &  Redis \\
\emph{person-controller} &  16 &  1112 &  12 &  &  MongoDB \\
\emph{proxyprint} &  73 &  8338 &  74 & \checkmark &  H2 \\
\emph{quartz-manager} &  129 &  5068 &  11 & \checkmark &   \\
\emph{reservations-api} &  39 &  1853 &  7 & \checkmark &  MongoDB \\
\emph{rest-ncs} &  9 &  605 &  6 &  &   \\
\emph{rest-news} &  11 &  857 &  7 &  &  H2 \\
\emph{rest-scs} &  13 &  862 &  11 &  &   \\
\emph{restcountries} &  24 &  1977 &  22 &  &   \\
\emph{scout-api} &  93 &  9736 &  49 & \checkmark &  H2 \\
\emph{session-service} &  15 &  1471 &  8 &  &  MongoDB \\
\emph{spring-actuator-demo} &  5 &  117 &  2 & \checkmark &   \\
\emph{spring-batch-rest} &  65 &  3668 &  5 &  &   \\
\emph{spring-ecommerce} &  58 &  2223 &  26 & \checkmark &  MongoDB;Redis;Elasticsearch \\
\emph{spring-rest-example} &  32 &  1426 &  9 &  &  MySQL \\
\emph{swagger-petstore} &  23 &  1631 &  19 &  &   \\
\emph{tiltaksgjennomforing} &  472 &  27316 &  79 & \checkmark &  PostgreSQL \\
\emph{tracking-system} &  87 &  5947 &  67 & \checkmark &  H2 \\
\emph{user-management} &  69 &  4274 &  21 &  &  MySQL \\
\emph{webgoat} &  355 &  27638 &  204 & \checkmark &  H2 \\
\emph{youtube-mock} &  29 &  3229 &  1 &  &   \\
\midrule
Total 36 & 6465 & 657162 & 1487 & 15 & 25 \\
\bottomrule 
\end{tabular} 

}
\end{table}

To answer {\bf RQ3}, we selected an existing corpus of REST APIs used in several studies in API fuzzing research, called WFD~\cite{sahin_2025_wfc}, formerly known as EMB~\cite{icst2023emb}, which has been available open-source and extended throughout the years since 2017.
We selected \emph{all} APIs in its most recent version.\footnote{\url{https://github.com/WebFuzzing/Dataset}}
Table~\ref{tab:suts-wb} shows some statistics on these 36 APIs.
WFD is a variegated corpus, including small example APIs, as well as large real-world APIs (e.g., coming from public administrations from different countries such as Germany, Norway, Thailand and the United Kingdom).
For the APIs that need it, authentication information in WFD is provided in YAML files, using the WFC format~\cite{sahin_2025_wfc}, like for the experiments for {\bf RQ2}.

Note that security faults can be found only if they are present.
Access policy violations cannot be found if an API has no authentication mechanism.
Likewise, no SQL Injection can be found if the API does not use any SQL database.
We have not injected any artificial fault in any of these APIs.
In our experiments, these 36 APIs are used as they are.

To answer {\bf RQ4}, we use the data from the experiments for % {\bf RQ2} and
{\bf RQ3}, where we measure the execution time of our security testing phase after the fuzzing phase.

In total, for our experiments in this paper we used $9+8+36-1=52$ distinct APIs, for a total of more than $800$ thousand lines of code (for business logic, not including third-party libraries), and more than $1600$ endpoints.
Note that \emph{webgoat} is used twice, both for {\bf RQ2} and {\bf RQ3}.
The API \emph{webgoat} is a deliberately insecure application, built and maintained by OWASP.
As such, it is part of {\bf RQ2}.
However, as it is written in Java, it was added to WFD in the past.
Therefore, we run it as well in {\bf RQ3}.
Note that, as discussed in the next section, experiments for  {\bf RQ2} and {\bf RQ3} are run differently (i.e., \emph{black-box} vs.~\emph{white-box} testing).
As such, reusing a single API twice should not be problematic.

%--------------------------------------------------------------
\subsection{Experiments Setup}

The 9 artificial examples used for answering {\bf RQ1} are small and simple.
As such, experiments were run with our extension of \evo for just 1 minute, 10 times, for a total of 90 minutes.

Experiments for {\bf RQ2} were run using \emph{black-box} mode of \evo.
On each of these 8 APIs, \evo was run for 1 hour.
Experiments were repeated 10 times per API, for a total of 80 hours.

As all the APIs used for {\bf RQ3} are written in Java or Kotlin, to answer this research question we use the \emph{white-box} mode
of \evo.
We use the same time budget and repetition settings as in {\bf RQ2}, i.e.,   \evo was run for 1 hour, with each experiments repeated 10 times.
This resulted in a duration of  360 hours.

Our novel oracles can work both for black-box and white-box testing.
We use different settings in {\bf RQ2} and {\bf RQ3} to show this being indeed the case.
In total, our experiments took more than 440 hours, i.e., more than 18 days if run in sequence.

Answering {\bf RQ4} did not require running any further experiment, as based on the same data for the other RQs.

%--------------------------------------------------------------
\subsection{Results for RQ1}

In every single experiment for {\bf RQ1}, the injected faults were correctly identified.
When there are no complex input constraints that prevent reaching the execution of the faulty code,
our novel oracles are able to correctly identify all the security vulnerabilities that we have manually injected.

\evo has an advanced and sophisticated system of end-to-end tests to verify its correctness~\cite{arcuri2023building}.
There are several artificial APIs in which \evo is automatically run on, and the generated tests are compiled and run, verifying different properties (e.g., if injected faults are properly identified).
These checks on these artificial APIs are run as JUnit tests as part of \evo's own continuous integration (CI) system~\cite{arcuri2023building} (e.g., on GitHub Actions).
These nine new artificial examples are now part of these regression test suites, executed in CI at each new code commit of \evo, as they can be  reliably handled by \evo.

\begin{result}
{\bf RQ1}: On simple APIs with injected faults, our techniques are able to reliably identify those security vulnerabilities.
\end{result}

%--------------------------------------------------------------
\subsection{Results for RQ2}

\begin{table}[!t]
\centering
\caption{ Detected faults on the 8 vulnerable APIs.
\label{tab:results-bb}
}
\vspace{-1.5\baselineskip}
% \begin{adjustbox}{width=.95\textwidth,center}
\begin{tabular}{lr|rrrrrr }\\ 
\toprule 
 SUT & \#Endp   & F100 & F201 & F204 & F205 & F902 & F903  \\ 
\midrule 
\emph{capital} & 21 &  &  & 1.0 &  &  &  \\ 
\emph{crapi} & 44 & 6.0 &  &  &  &  & 3.0 \\ 
\emph{damn-vulnerable-rest} & 21 & 5.3 & 1.0 & 0.3 & 1.2 &  &  \\ 
\emph{dvapi} & 16 & 3.0 &  &  &  &  & 10.0 \\ 
\emph{dvws-node} & 31 & 4.0 & 4.0 &  &  &  &  \\ 
\emph{vampi} & 14 & 2.2 & 0.1 &  &  & 0.7 & 5.0 \\ 
\emph{vulnerable-rest-api} & 19 & 5.0 &  &  &  &  &  \\ 
\emph{webgoat} & 201 & 31.0 & 12.4 &  & 5.0 & 28.0 & 3.0 \\ 
\midrule 
Mean  & 46 & 7.1 & 2.2 & 0.2 & 0.8 & 3.6 & 2.6 \\ 
Median  & 21 & 4.5 & 0.1 & 0.0 & 0.0 & 0.0 & 1.5 \\ 
Sum  & 367 & 56.5 & 17.5 & 1.3 & 6.2 & 28.7 & 21.0 \\ 
\bottomrule 
\end{tabular} 

% \end{adjustbox}
\end{table}

% \begin{table*}[!t]
% \centering
% \caption{\label{tab:results}
% Detected faults, per API, averaged by number of runs (10).
% We report faults based on the existing oracles in \evo,
% including checking for returned HTTP 500 status codes (\emph{H500}),
% and faults related to mismatched responses based on what declared in the OpenAPI schemas (\emph{Schema}).
% Results for our three novel oracles are labelled as
% \emph{Not-Recognized} (Section~\ref{sec:not-recognized}),
% \emph{Leakage} (Section~\ref{sec:leakage})
% and
% \emph{Missed-Check}    (Section~\ref{sec:missed-checks}).
% }
% \vspace{-1.5\baselineskip}
% \input{generated_files/tableCodes.tex}
% \end{table*}

Table~\ref{tab:results-bb} shows the results  for  {\bf RQ2} on eight vulnerable APIs.
Besides the results for our novel oracles, we also report the number of faults related to HTTP 500 status codes,
which is the most common oracle in the REST API fuzzing literature.
\evo can also detect other types of functional faults (e.g., schema mismatches) and SSRF, but those are not discussed and presented in this paper.
Faults are based on the endpoint calls that led to trigger them.
For example, if an API has $n$ endpoints, there could be detected at most $n$ distinct faults related to HTTP 500.
It could be possible that the code of an endpoint has more than one fault that could be triggered differently, or that the same one single fault can lead to more than one endpoint to fail.
Counting based on number of failing endpoints is a common, reasonable compromise when counting the number of faults, especially in black-box testing, where a root-cause analysis cannot be easily automated.

To simplify the discussion and references to these faults, each one has a unique label, based on the labeling system defined in WFC~\cite{sahin_2025_wfc}.
In particular, we have:
\begin{itemize}
\item \texttt{F100}: HTTP Status 500
\item \texttt{F200}: SQL Injection (Section~\ref{sec:F200})
\item \texttt{F201}: Cross Site Scripting (XSS) (Section~\ref{sec:F201})
\item \texttt{F204}: Existence Leakage (Section~\ref{sec:F204})
\item \texttt{F205}: Not Recognized Authentication (Section~\ref{sec:F205})
\item \texttt{F206}: Missed Authorization Checks (Section~\ref{sec:F206})
\item \texttt{F900}: Ignore Anonymous (Section~\ref{sec:F900})
\item \texttt{F901}: Anonymous Modifications (Section~\ref{sec:F901})
\item \texttt{F902}: Leaked Stack Trace (Section~\ref{sec:F902})
\item \texttt{F903}: Hidden Accessible (Section~\ref{sec:F903})
\end{itemize}

Note that codes \texttt{F1xx} are for functional oracles, whereas \texttt{F2xx} are for security faults.
Codes in the group \texttt{9xx} are for work-in-progress codes not part of WFC yet~\cite{sahin_2025_wfc}.
The use of unique, stable codes is meant for simplifying automated comparisons of fuzzers that use the WFC convention.
Table~\ref{tab:results-bb} shows only data for which faults were found.
If a fault type was never found in any of the 10 runs on any of the 8 APIs, such data is not present in the table (to avoid cluttering tables with 0-value columns).

In these experiments, tens of security faults were found, but only of five different kinds.
No fault for
\texttt{F200} (SQL Injection),
\texttt{F206} (Missed Authorization Checks),
\texttt{F900} (Ignore Anonymous)
and
\texttt{F901} (Anonymous Modifications)
were found.
Faults can only be found if they exist.
We have not manually reviewed the whole code and documentation of those 8 APIs to see if there are any other faults that could had been found
with our oracles, or if what found is all that is there.
Still, the fact that our novel techniques could generate test cases that can detect these faults show that they can be of practical value.

Looking at the generated tests and logs of these experiments, one issue was clear though.
Some of these vulnerable-by-design example APIs (e.g., \emph{crapi}, \emph{capital} and \emph{vampi}) have endpoints that allow modifying login information about the users.
This is currently a major issue for fuzzers in \emph{black-box} testing, as such data can be modified, making the login credentials provided at the beginning of fuzzing no longer valid.
This could explain why some faults were not found.
However, this is a not a problem in \emph{white-box} testing, as database state can be automatically and efficiently reset after each test execution (e.g.,~\cite{arcuri2020sql}).
Also, for real-world APIs, typically login credentials are handled and managed in separated, ad-hoc APIs.

An alternative would be to run experiments by excluding these endpoints.
For example, that is what we did previously in~\cite{arcuri2025fuzzing},
where it was then possible to detect a
\texttt{F206} (Missed Authorization Checks)
in \emph{capital}.
However, it would be better to design some automated system to handle this issue.

\begin{result}
{\bf RQ2}: Our novel oracles were effective at detecting tens of faults in vulnerable APIs used as security examples.
\end{result}

%--------------------------------------------------------------
\subsection{Results for RQ3}

\begin{table}[!t]
\centering
\caption{ Detected faults on the 36 APIs from WFD.
\label{tab:results-wb}
}
\vspace{-1.5\baselineskip}
\begin{adjustbox}{width=.95\textwidth,center}
\begin{tabular}{lr|rrrrrrrr }\\ 
\toprule 
 SUT & \#Endp   & F100 & F201 & F204 & F205 & F900 & F901 & F902 & F903  \\ 
\midrule 
\emph{bibliothek} & 8 & 2.7 &  &  &  &  &  &  & 26.0 \\ 
\emph{blogapi} & 52 & 28.5 & 0.1 &  & 0.5 &  &  &  & 7.0 \\ 
\emph{catwatch} & 14 & 6.4 &  &  &  & 6.2 &  & 2.5 &  \\ 
\emph{cwa-verification} & 5 & 5.0 &  &  &  &  &  &  &  \\ 
\emph{erc20-rest-service} & 13 & 12.0 &  &  &  &  &  &  &  \\ 
\emph{familie-ba-sak} & 183 & 151.2 & 2.0 &  & 12.0 & 4.6 &  &  &  \\ 
\emph{features-service} & 18 & 12.5 &  &  &  &  & 1.0 & 12.3 &  \\ 
\emph{genome-nexus} & 23 &  & 5.0 &  &  &  &  &  & 2.0 \\ 
\emph{gestaohospital} & 20 & 1.9 & 2.2 &  &  &  & 3.6 &  & 6.0 \\ 
\emph{http-patch-spring} & 6 &  & 0.5 &  &  &  & 1.2 &  &  \\ 
\emph{languagetool} & 2 & 1.0 &  &  &  &  &  &  &  \\ 
\emph{market} & 13 & 5.3 &  &  &  &  &  &  &  \\ 
\emph{microcks} & 88 & 10.2 & 7.9 &  &  &  &  &  & 14.0 \\ 
\emph{ocvn} & 258 & 194.3 &  &  &  &  &  & 82.9 &  \\ 
\emph{ohsome-api} & 134 & 1.6 &  &  &  &  &  &  &  \\ 
\emph{pay-publicapi} & 10 & 10.0 &  &  &  &  &  &  &  \\ 
\emph{person-controller} & 12 & 11.0 &  &  &  &  &  & 5.0 & 2.0 \\ 
\emph{proxyprint} & 74 & 33.9 &  & 2.0 & 3.7 &  &  & 8.7 &  \\ 
\emph{quartz-manager} & 11 & 5.9 &  &  &  &  &  &  &  \\ 
\emph{reservations-api} & 7 & 3.8 &  &  & 0.1 &  &  &  & 2.0 \\ 
\emph{rest-ncs} & 6 &  &  &  &  &  &  &  &  \\ 
\emph{rest-news} & 7 & 2.9 &  &  &  &  & 2.4 &  &  \\ 
\emph{rest-scs} & 11 &  &  &  &  &  &  &  &  \\ 
\emph{restcountries} & 22 & 1.0 &  &  &  &  &  &  &  \\ 
\emph{scout-api} & 49 & 15.2 & 0.3 &  &  &  &  &  &  \\ 
\emph{session-service} & 8 & 3.6 &  &  &  &  &  &  & 2.0 \\ 
\emph{spring-actuator-demo} & 2 &  & 1.0 &  &  &  &  &  &  \\ 
\emph{spring-batch-rest} & 5 & 1.8 & 1.0 &  &  &  &  &  &  \\ 
\emph{spring-ecommerce} & 27 & 15.8 &  &  & 3.1 & 0.3 &  &  &  \\ 
\emph{spring-rest-example} & 9 & 1.0 & 0.4 &  &  &  & 4.0 &  &  \\ 
\emph{swagger-petstore} & 19 & 9.2 & 3.8 &  &  &  & 3.0 &  &  \\ 
\emph{tiltaksgjennomforing} & 79 & 28.7 & 1.0 &  &  &  &  &  & 8.0 \\ 
\emph{tracking-system} & 67 &  & 10.6 &  &  &  & 1.9 &  & 21.0 \\ 
\emph{user-management} & 21 & 6.7 & 5.2 &  &  &  & 3.7 &  & 4.0 \\ 
\emph{webgoat} & 204 & 5.0 &  &  &  &  &  & 4.0 & 2.0 \\ 
\emph{youtube-mock} & 1 &  &  &  &  &  &  &  &  \\ 
\midrule 
Mean  & 41 & 16.3 & 1.1 & 0.1 & 0.5 & 0.3 & 0.6 & 3.2 & 2.7 \\ 
Median  & 14 & 5.0 & 0.0 & 0.0 & 0.0 & 0.0 & 0.0 & 0.0 & 0.0 \\ 
Sum  & 1488 & 588.1 & 41.0 & 2.0 & 19.4 & 11.1 & 20.8 & 115.4 & 96.0 \\ 
\bottomrule 
\end{tabular} 

\end{adjustbox}
\end{table}

When it comes to white-box testing of the APIs in WFD, in Table~\ref{tab:results-wb} we can see that security faults are found in most of the APIs, with hundreds of distinct cases.
However, no fault of types
\texttt{F200} (SQL Injection)
and
\texttt{F206} (Missed Authorization Checks)
were found.
These are among the most critical types of faults that we can detect with our novel techniques.

Table~\ref{tab:results-wb} shows that, on the 36 analyzed APIs, in total our oracles detected more than 300 distinct faults (although not all were found in all runs). %, recall the discussion about \emph{scout-api} for example).
This provides evidence for the practical value of our novel techniques.

% /**
% * Calls:
% * 1 - (401) GET:/api/task/logg/{id}
% * 2 - (200) GET:/api/task/logg/{id}
% * 3 - (401) POST:/api/ekstern/pensjon/hent-barnetrygd
% * Found 1 potential fault of type-code 205
% * Using 1 example:
% *   (-) BODY -> 2020-12-01
% */

\begin{figure*}
\begin{lstlisting}[language=Java,numbers=left,xleftmargin=2em,basicstyle=\footnotesize\ttfamily]
@Test @Timeout(60)
public void test_138_postOnHent_barnetrygdAuthenticatedButWronglyToldNot() throws Exception {

    final String token_Veileder = given()
            .contentType("application/x-www-form-urlencoded")
            .body("name=Veileder&grant_type=client_credentials&code=foo&client_id=foo&client_secret=secret")
            .post("http://localhost:10512/azuread/token")
            .then().extract().response().path("access_token"); ©\label{line:auth0}©

    final String auth_Veileder = "Bearer " + token_Veileder;
    final String token_TaskRunner = given()
            .contentType("application/x-www-form-urlencoded")
            .body("name=TaskRunner&grant_type=client_credentials&code=foo&client_id=foo&client_secret=secret")
            .post("http://localhost:10512/azuread/token")
            .then().extract().response().path("access_token"); ©\label{line:auth1}©

    final String auth_TaskRunner = "Bearer " + token_TaskRunner;

    given().accept("*/*")
            .header("x-EMextraHeader123", "_EM_14362_XYZ_")
            .header("Authorization", auth_Veileder) // Veileder
            .get(baseUrlOfSut + "/api/task/logg/710?page=352")
            .then()
            .statusCode(401) ©\label{line:first}©
            .assertThat()
            .contentType("text/html");

    given().accept("*/*")
            .header("x-EMextraHeader123", "42")
            .header("Authorization", auth_TaskRunner) // TaskRunner
            .get(baseUrlOfSut + "/api/task/logg/718")
            .then()
            .statusCode(200) ©\label{line:second}©
            .assertThat()
            .contentType("application/json")
            .body("'data'", nullValue())
            .body("'status'", containsString("FEILET"))
            .body("'melding'", containsString("Henting av tasklogg feilet."))
            .body("'frontendFeilmelding'", containsString("Henting av tasklogg feilet."));

    // Fault205. Wrongly Not Recognized as Authenticated.
    given().accept("application/json")
            .header("x-EMextraHeader123", "42")
            .header("Authorization", auth_TaskRunner) // TaskRunner
            .contentType("application/json")
            .body(" { " +
                " \"ident\": \"_EM_14364_XYZ_\", " +
                " \"fraDato\": \"2020-12-01\" " +
                " } ")
            .post(baseUrlOfSut + "/api/ekstern/pensjon/hent-barnetrygd?EMextraParam123=_EM_14366_XYZ_")
            .then()
            .statusCode(401) ©\label{line:third}©
            .assertThat()
            .contentType("application/json")
            .body("'servlet'", containsString("dispatcherServlet"))
            .body("'message'", containsString("Pensjon tjeneste kan kun kalles av pensjon eller innlogget bruker med FORVALTER rolle")) ©\label{line:message}©
            .body("'url'", containsString("/api/ekstern/pensjon/hent-barnetrygd"))
            .body("'status'", containsString("401"));
}
\end{lstlisting}
\caption{\label{fig:nav}
Snippet example of generated test case in which a \emph{Not Recognized Authentication} (Section~\ref{sec:F205}) fault is
identified in \emph{familie-ba-sak}.
Due to reason of space, the generated test's JavaDoc is not shown.
}
\end{figure*}

An API like \emph{familie-ba-sak} is large and complex (Table~\ref{tab:suts-wb}).
It is made by the public administration in Norway,
 dealing with ``child benefit case processing'', used by millions of people.\footnote{\url{https://github.com/navikt/familie-ba-sak}}
Finding a Missed-Check or SQL Injection would had rather been unsettling, and we are glad none was found.
(If any was found, we would have contacted the Norwegian authorities before making this paper public).

Interestingly, there were several cases of \emph{Not Recognized Authentication} (Section~\ref{sec:F205}).
Figure~\ref{fig:nav} shows an example of generated test detecting a \emph{Not Recognized Authentication} (Section~\ref{sec:F205}) fault.
This automatically generated test case starts with 2 calls to the authentication service, to get dynamic authentication tokens
for the users \texttt{Veileder} (Line~\ref{line:auth0}) and \texttt{TaskRunner} (Line~\ref{line:auth1}).
The first call on \texttt{/api/task/logg/710} with \texttt{Veileder} shows that such endpoint requires authentication/authorization,
as a 401 status code is returned (Line~\ref{line:first}).
A second call on the same endpoint with \texttt{TaskRunner} gets a 200 status code (Line~\ref{line:second}).
This provides evidence that the credentials for \texttt{TaskRunner} are not wrongly set.
However, a call with same user on \texttt{/api/ekstern/pensjon/hent-barnetrygd} returns a 401 (Line~\ref{line:third}), which would imply the user
is not recognized.
Based on previous calls, this is wrong.
Note that, even if access tokens might have a short life-span, they are dynamically retrieved in the same test case, and, as such, after a few
(milli)seconds they should still be valid.
Interestingly, the error message (in Norwegian at Line~\ref{line:message}) seems to support the fact that this is a bug (i.e., a 403 should had been returned),
as it states, translated with Google Translate: ``Pension service can only be called by pension or logged in user with ADMINISTRATOR role''.

Not all faults have the same severity.
For example, a 500 status does not really mean that the API crashed.
If it crashed, then it would not be able to send back a HTTP response with status 500, and the TCP socket would had been closed.
Typically, when an exception is thrown,  then the framework (e.g., Spring for Java) running the API catches
the exception, and sends back an error response.
The API will keep running.
If the crash is due to invalid input, then the fault is related to improper input validation, and rather a 4xx status error (indicating an
 user error) should had been returned.
 A typical example is assuming a requested resource to exist, and crashing on a null-pointer exception instead of returning a 404
 not found user error status~\cite{marculescu2022faults}.
This is still a fault that needs to be fixed, but it is not as critical as it might look at a first sight.
In this paper, we have proposed 7 novel oracles, but we cannot say how critical they are for certain,
unless empirical studies or interviews/surveys
with practitioners were carried out to assess their assumed importance.
However, we argue that a security vulnerability that leads to unauthorized access of resources (e.g., \emph{Missed-Check})
is likely much more important than wrongly returning a 500 status code instead of a 404.

In the data generated in our experiments, there are hundreds of interesting cases that could be worth discussing.
For reason of space, we cannot discuss all of them.
However, we noticed that, in a few cases, when our novel oracles detected faults, the problem was not due to a security vulnerability, but rather due to severe faults related to HTTP semantics.
To clarify, we will discuss here some of these cases in more details.

%-----------------------------------------------------------------
\begin{figure}
\begin{lstlisting}[language=Java,numbers=left,xleftmargin=2em]
/**
* Calls:
* 1 - (200) OPTIONS:/api/users/checkEmailAvailability
* 2 - (415) PUT:/api/users/checkEmailAvailability
* Found 3 potential faults. Type-codes: 101, 903
*/
@Test(timeout = 60000)
public void test_46_putOnCheckEmailAvailabilityShowsFaults_101_903() throws Exception {

    final String token_admin = given()
            .contentType("application/json")
            .body(" { " +
                " \"usernameOrEmail\": \"admin\", " +
                " \"password\": \"bar123\" " +
                " } ")
            .post(baseUrlOfSut + "/api/auth/signin")
            .then().extract().response().path("accessToken");

    final String auth_admin = "Bearer " + token_admin;

    // Fault101. Received A Response From API With A Structure/Data That Is Not Matching Its Schema. Type: validation.request.operation.notAllowed OPTIONS operation not allowed on path '/api/users/checkEmailAvailability'.
    given().accept("*/*")
            .header("Authorization", auth_admin) // admin
            .options(baseUrlOfSut + "/api/users/checkEmailAvailability")
            .then()
            .statusCode(200)
            .assertThat()
            // .header("Allow", "HEAD,DELETE,GET,OPTIONS,PUT")
            .body(isEmptyOrNullString());

    // Fault101. Received A Response From API With A Structure/Data That Is Not Matching Its Schema. Type: validation.response.body.unexpected No response body is expected but one was found.
    // Fault903. Hidden Accessible Endpoint.
    given().accept("*/*")
            .header("Authorization", auth_admin) // admin
            .put(baseUrlOfSut + "/api/users/checkEmailAvailability")
            .then()
            .statusCode(415)
            .assertThat()
            .contentType("application/json")
            .body("'status'", numberMatches(415))
            .body("'error'", containsString("Unsupported Media Type"))
            .body("'message'", containsString("Content type 'application/x-www-form-urlencoded;charset=UTF-8' not supported"))
            .body("'path'", containsString("/api/users/checkEmailAvailability"));
}
\end{lstlisting}
\caption{\label{fig:903:blogapi}
Example of generated test for \emph{blogapi} finding a \texttt{F903} \emph{Hidden Accessible Endpoint} fault.
}
\end{figure}

%-----------------------------------------------------------------
\begin{figure}
\begin{lstlisting}[language=Java,numbers=left,xleftmargin=2em]
/**
* Calls:
* 1 - (404) OPTIONS:/v2/projects/{project}
* 2 - (404) POST:/v2/projects/{project}
* Found 1 potential fault of type-code 903
*/
@Test @Timeout(60)
public void test_6_postOnProjectHiddenAccessible() throws Exception {

    given().accept("*/*")
            .options(baseUrlOfSut + "/v2/projects/omm_postfix")
            .then()
            .statusCode(404)
            .assertThat()
            // .header("Allow", "HEAD,DELETE,POST,GET,OPTIONS,PUT,PATCH")
            .contentType("application/json")
            .body("'error'", containsString("Endpoint not found."));

    // Fault903. Hidden Accessible Endpoint.
    given().accept("*/*")
            .post(baseUrlOfSut + "/v2/projects/omm_postfix")
            .then()
            .statusCode(404)
            .assertThat()
            .contentType("application/json")
            .body("'error'", containsString("Endpoint not found."));
}
\end{lstlisting}
\caption{\label{fig:903:bibliotek}
Example of generated test for \emph{bibliotek} finding a \texttt{F903} \emph{Hidden Accessible Endpoint} fault.
}
\end{figure}

Figure~\ref{fig:903:blogapi} and Figure~\ref{fig:903:bibliotek}
show two distinct examples of \texttt{F903} \emph{Hidden Accessible Endpoint} faults,
in the \emph{blogapi} and \emph{bibliotek} APIs.
After a manual investigation of these APIs, there is no hidden accessible endpoint in these APIs.
There are two other problems though.
First, the \texttt{OPTIONS} is returning what could be consider as ``garbage'', claiming in the \emph{Allow} header the presence of endpoints that do not exist.
Second, calling those endpoints does not return a 405 (or 501), as expected by HTTP specifications.
They  rather return a 415 (for \emph{blogapi}) or 404 (\emph{bibliotek}).
Either case, these response statuses make no sense according the semantics of HTTP.
This is not a fault in the implementation of the business logic of these APIs, but a rather a serious fault or misconfiguration in their employed HTTP frameworks.

%-----------------------------------------------------------------
\begin{figure}
\begin{lstlisting}[language=Java,numbers=left,xleftmargin=2em,basicstyle=\footnotesize\ttfamily]
/**
* Calls:
* 1 - (403) GET:/config
* 2 - (200) GET:/config
* Found 3 potential faults. Type-codes: 101, 900
*/
@Test(timeout = 60000)
public void test_69_getOnConfigShowsFaults_101_900() throws Exception {

        // Fault101. Received A Response From API With A Structure/Data That Is Not Matching Its Schema. Type: validation.response.body.unexpected No response body is expected but one was found.
        given().accept("application/json")
                        .header("x-EMextraHeader123", "")
                        .header("X-Organizations", "urLJpDVv")
                        .get(baseUrlOfSut + "/config?" +
                                "access_token=cvOYB8K5cuf&" +  ©\label{line:catwatch}©
                                "q=HXDhFkqk")
                        .then()
                        .statusCode(403)
                        .assertThat()
                        .contentType("application/json")
                        .body("'error'", containsString("access_denied"))
                        .body("'error_description'", containsString("Unable to obtain a new access token for resource &#39;null&#39;. The provider manager is not configured to support it."));

        // Fault101. Received A Response From API With A Structure/Data That Is Not Matching Its Schema. Type: validation.response.body.schema.type [Path '/github.login'] Instance type (null) does not match any allowed primitive type (allowed: ["string"])
        // Fault900. A Protected Resource Is Accessible Without Providing Any Authentication.
        given().accept("application/json")
                        .header("x-EMextraHeader123", "")
                        .get(baseUrlOfSut + "/config?" +
                                "q=_EM_6668_XYZ_&" +
                                "organizations=LVszttcjhEwqOL1&" +
                                "limit=tawdhQdONfwwEwo")
                        .then()
                        .statusCode(200)
                        .assertThat()
                        .contentType("application/json")
                        .body("'cache.path'", containsString("D:\\WORK\\EXPERIMENTS\\2026-security\\wb_0_9/temp/tmp_catwatch/cache_10212"))
                        .body("'cache.size'", containsString("50"))
                        .body("'endpoints.enabled'", containsString("false"))
                        .body("'endpoints.health.enabled'", containsString("true"))
                        .body("'github.login'", nullValue())
                        .body("'organization.list'", containsString("zalando,zalando-stups,zalando-incubator"))
                        .body("'schedule'", containsString("0 1 8 * * *"))
                        .body("'scoring.project'", containsString("function(project) {var daysSinceLastPush = 0;if (project.lastPushed) {var tokens = project.lastPushed.split(\" \");var day = tokens[2];var month = tokens[1];var year = tokens[5];var lastPushedDate = new Date(day + ' ' + month + ' ' + year);var millisInDay = 86400000;daysSinceLastPush = Math.floor((new Date() - lastPushedDate) / millisInDay);}var maintainersPenalty = 0;if (project.maintainers.length < 2) {maintainersPenalty = 100;}return project.starsCount * 3 + project.forksCount * 2 + project.contributorsCount * 5 - daysSinceLastPush - maintainersPenalty}"))
                        .body("'spring.database.driverClassName'", nullValue())
                        .body("'spring.datasource.platform'", nullValue())
                        .body("'spring.datasource.username'", containsString("sa"))
                        .body("'spring.jpa.database'", nullValue())
                        .body("'spring.jpa.hibernate.ddl-auto'", containsString("update"));
}
\end{lstlisting}
\caption{\label{fig:900:catwatch}
Example of generated test for \emph{catwatch} finding a \texttt{F900} \emph{Ignore Anonymous} fault.
}
\end{figure}

%-----------------------------------------------------------------
% \begin{figure}
% \begin{lstlisting}[language=Java,numbers=left,xleftmargin=2em,basicstyle=\tiny\ttfamily]
% /**
% * Calls:
% * 1 - (200) POST:/testverktoy/vedtak-om-overgangsstønad
% * 2 - (403) POST:/testverktoy/vedtak-om-overgangsstønad
% * 3 - (200) POST:/testverktoy/vedtak-om-overgangsstønad
% * Found 1 potential fault of type-code 900
% */
% @Test @Timeout(60)
% public void test_9_postOnVedtak_om_overgangsst_nadIgnoreAnonymous() throws Exception {
%
%         final String token_Veileder = given()
%                         .contentType("application/x-www-form-urlencoded")
%                         .body("name=Veileder&grant_type=client_credentials&code=foo&client_id=foo&client_secret=secret")
%                         .post("http://localhost:10512/azuread/token")
%                         .then().extract().response().path("access_token");
%
%         final String auth_Veileder = "Bearer " + token_Veileder;
%         final String token_Forvalter = given()
%                         .contentType("application/x-www-form-urlencoded")
%                         .body("name=Forvalter&grant_type=client_credentials&code=foo&client_id=foo&client_secret=secret")
%                         .post("http://localhost:10512/azuread/token")
%                         .then().extract().response().path("access_token");
%
%         final String auth_Forvalter = "Bearer " + token_Forvalter;
%
\begin{figure}
\begin{lstlisting}[language=Java,numbers=left,xleftmargin=2em,basicstyle=\footnotesize\ttfamily]
        ...
        given().accept("*/*")
                        .header("x-EMextraHeader123", "")
                        .header("Authorization", auth_Veileder) // Veileder
                        .contentType("application/json")
                        .body(" { " +
                                " \"ident\": \"\" " +
                                " } ")
                        .post(baseUrlOfSut + "/testverktoy/vedtak-om-overgangsstønad")
                        .then()
                        .statusCode(200) ©\label{line:900:veileder}©
                        .assertThat()
                        .contentType("application/json")
                        .body("'data'", nullValue())
                        .body("'status'", containsString("FEILET"))
                        .body("'melding'", containsString("Fant ikke aktiv ident for aktør"))
                        .body("'frontendFeilmelding'", nullValue())
                        .body("'stacktrace'", nullValue());

        given().accept("*/*")
                        .header("x-EMextraHeader123", "")
                        .header("Authorization", auth_Forvalter) // Forvalter
                        .contentType("application/json")
                        .body(" { " +
                                " \"ident\": \"\" " +
                                " } ")
                        .post(baseUrlOfSut + "/testverktoy/vedtak-om-overgangsstønad")
                        .then()
                        .statusCode(403) ©\label{line:900:forvalter}©
                        .assertThat()
                        .contentType("text/html");

        // Fault900. A Protected Resource Is Accessible Without Providing Any Authentication.
        ValidatableResponse res_2 = given().accept("*/*")
                        .header("x-EMextraHeader123", "")
                        .contentType("application/json")
                        .body(" { " +
                                " \"ident\": \"\" " +
                                " } ")
                        .post(baseUrlOfSut + "/testverktoy/vedtak-om-overgangsstønad")
                        .then()
                        .statusCode(200)  ©\label{line:900:anonymous}©
                        .assertThat()
                        .contentType("application/json")
                        .body("'data'", nullValue())
                        .body("'status'", containsString("FEILET"))
                        .body("'melding'", containsString("Fant ikke aktiv ident for aktør"))
                        .body("'frontendFeilmelding'", nullValue())
                        .body("'stacktrace'", nullValue());
        String location_vedtak_om_overgangsst_nad__98 = res_2.extract().header("location");
        assertTrue(isValidURIorEmpty(location_vedtak_om_overgangsst_nad__98));
}
\end{lstlisting}
\caption{\label{fig:900:familie-ba-sak}
Snippet example of generated test for \emph{familie-ba-sak} finding a \texttt{F900} \emph{Ignore Anonymous} fault.
For reason of space, the calls retrieving the authentication tokens are omitted.
}
\end{figure}

Figure~\ref{fig:900:catwatch} and Figure~\ref{fig:900:familie-ba-sak}
show two distinct examples of \texttt{F900} \emph{Ignore Anonymous} faults,
in the \emph{catwatch} and \emph{familie-ba-sak} APIs.
In theory, this is a potentially catastrophic vulnerability.
The API \emph{catwatch} has no defined authentication information in WFD (recall Table~\ref{tab:suts-wb}).
A call to the endpoint \texttt{GET:/config} without any authentication info returns a 200 with some data.
However, a non-authenticated call with the query parameter \texttt{access\_token=cvOYB8K5cuf} (Line~\ref{line:catwatch})
leads to return a 403.
Without an in-depth analysis of the semantics of this API, it is unclear if this is a major security vulnerability, or misuse of HTTP status codes.
Either way, it is problematic.

The case of \emph{familie-ba-sak} in Figure~\ref{fig:900:familie-ba-sak} is easier to analyze, but potentially more disturbing.
An authenticated user ``Veileder'' makes a \texttt{POST} to an endpoint, successfully returning a 200 status code (Line~\ref{line:900:veileder}).
A different user  ``Forvalter'', with different permission rights, is blocked from doing the same operation, and obtains a 403 (Line~\ref{line:900:forvalter}).
Finally, a call without any authentication information is successful with a 200 status code (Line~\ref{line:900:anonymous}).
Without looking at the response, this could be a major security issue.
However, \texttt{FEILET} in Norwegian means ``the error'', and ``\texttt{Fant ikke aktiv ident for aktør}'' means ``No active ident found for actor''.
In other words, the call fails, but the API decides to return a 200 success status code, which seems completely wrong in this context.

The case of Figure~\ref{fig:900:familie-ba-sak} is interesting because, without a natural-language processing (NLP) of the text of the responses, it could be difficult to automatically detect when an API is wrongly returning a 2xx on a failed call.
And, even if NLP techniques are used, there is the need to make sure they work for different human languages (e.g., Norwegian Bokmål or Nynorsk, as in this case).
Several of our security oracles rely on the correct use of the HTTP protocol (e.g., user errors should return a status code in the 4xx family).
When they detect a fault, they could indeed detect a security fault, or, like in this case, they might detect a major issue in the use of HTTP semantics in the tested API.
Either case, these are serious faults that need fixing.

\begin{result}
{\bf RQ3}: Our novel security oracles were effective at finding new faults in existing real-world APIs.
They either detect security vulnerabilities, or serious issues in the use of the HTTP protocol.
\end{result}

%--------------------------------------------------------------
\subsection{Results for RQ4}
\label{sec:rq4}

%-----------------------------------------------------------
\begin{table}[!t]
\centering
\caption{ For the 36 APIs in WFD, we report the obtained 2xx coverage of their endpoints, and the overhead in seconds of our new security phase.
\label{tab:overhead}
}
\vspace{-1.5\baselineskip}
%\begin{adjustbox}{width=.95\textwidth,center}
\begin{tabular}{ l rrr}\\ 
\toprule 
SUT & \#Endpoints & \% 2xx Coverage & Overhead \\ 
\midrule 
\emph{bibliothek} &  8 &  20.0  &  1.2  \\ 
\emph{blogapi} &  52 &  30.4  &  123.7  \\ 
\emph{catwatch} &  14 &  49.3  &  110.2  \\ 
\emph{cwa-verification} &  5 &  0.0  &  0.0  \\ 
\emph{erc20-rest-service} &  13 &  7.7  &  0.4  \\ 
\emph{familie-ba-sak} &  183 &  49.5  &  818.2  \\ 
\emph{features-service} &  18 &  21.1  &  5.4  \\ 
\emph{genome-nexus} &  23 &  72.9  &  37.1  \\ 
\emph{gestaohospital} &  20 &  58.3  &  12.4  \\ 
\emph{http-patch-spring} &  6 &  73.3  &  8.2  \\ 
\emph{languagetool} &  2 &  92.9  &  49.7  \\ 
\emph{market} &  13 &  43.1  &  58.3  \\ 
\emph{microcks} &  88 &  46.3  &  81.9  \\ 
\emph{ocvn} &  258 &  84.0  &  1986.5  \\ 
\emph{ohsome-api} &  134 &  2.8  &  14.8  \\ 
\emph{pay-publicapi} &  10 &  57.5  &  5.1  \\ 
\emph{person-controller} &  12 &  43.5  &  3.0  \\ 
\emph{proxyprint} &  74 &  67.7  &  2325.6  \\ 
\emph{quartz-manager} &  11 &  36.4  &  7.3  \\ 
\emph{reservations-api} &  7 &  57.1  &  91.5  \\ 
\emph{rest-ncs} &  6 &  100.0  &  1.0  \\ 
\emph{rest-news} &  7 &  91.4  &  9.2  \\ 
\emph{rest-scs} &  11 &  100.0  &  1.0  \\ 
\emph{restcountries} &  22 &  100.0  &  6.4  \\ 
\emph{scout-api} &  49 &  34.1  &  21.8  \\ 
\emph{session-service} &  8 &  65.3  &  10.1  \\ 
\emph{spring-actuator-demo} &  2 &  100.0  &  21.7  \\ 
\emph{spring-batch-rest} &  5 &  100.0  &  1.0  \\ 
\emph{spring-ecommerce} &  27 &  47.7  &  156.9  \\ 
\emph{spring-rest-example} &  9 &  92.6  &  35.0  \\ 
\emph{swagger-petstore} &  19 &  69.0  &  2.4  \\ 
\emph{tiltaksgjennomforing} &  79 &  8.9  &  8.7  \\ 
\emph{tracking-system} &  67 &  38.7  &  142.8  \\ 
\emph{user-management} &  21 &  66.7  &  61.2  \\ 
\emph{webgoat} &  204 &  16.7  &  54.3  \\ 
\emph{youtube-mock} &  1 &  22.2  &  0.0  \\ 
\midrule 
Average  &  &  54.6  &  174.3  \\ 
Median  &  &  53.3  &  13.6  \\ 
\bottomrule 
\end{tabular} 

%\end{adjustbox}
\end{table}

In fuzzing, there is a balance to strike between the time spent in generating high coverage test cases (either code coverage and/or REST API schema coverage) and the time spent in crafting ad-hoc test cases targeted at specific security vulnerabilities.
As explained in Section~\ref{sec:integration}, our security-phase is run as a post-processing, after the main fuzzing is completed.
In our context, the number of generated tests evaluated for security is fixed, depending on the number of endpoints and the type and quality of the generated tests during the initial fuzzing.
If the main fuzzing session is run for 1 hour, it is important to evaluate what is the additional overhead introduced by our security phase.

Table~\ref{tab:overhead} shows, for each API in WFD used for {\bf RQ3}, the overhead in seconds of our additional security phase.
To better analyze those results, we also report the 2xx schema coverage (i.e., the percentage of endpoints for which successful 2xx calls were created).
If no success call is made, and the execution of test cases ends at the first layer of input validation with a 4xx response, then it would be unlikely that any security testing would be successful, as most of our oracles require the ability to generate success calls to build upon.
Fuzzing REST APIs has witnessed significant improvements in the research literature, but it still has many open research problems that need addressing~\cite{zhang2023open}.

As shown in Table~\ref{tab:overhead}, in most cases our security phase is fast, taking few tens of seconds, or up to 2 minutes, which is a reasonable and practical amount of time considering a budget of 1-hour for the initial fuzzing session.
On these APIs, the median time is less than 14 seconds, whereas the average is less than 3 minutes.
Out of 36 APIs in WFD, there are, though, three outliers: \emph{familie-ba-sak} (13 minutes), \emph{ocvn} (33 minutes) and \emph{proxyprint} (38 minutes).
From a usability and practical standpoint, in the next versions of \evo it will be important to provide some further parameters to put timeouts on the security-phase, either in absolute terms (e.g., 10 minutes), or as proportion of the search budget (e.g., 20\% of the budget, which would be 12 minutes if the budget is 1 hour).

\begin{result}
{\bf RQ4}: In the large majority of cases, our novel techniques have small, negligible computational overhead.
\end{result}

%%%%%%%%%%%%%%%%%%%%%%%%%%%%%%%%%%%%%%%%%%%%%%%%%%%%%%%%%%%%%%%%%%%%%%%%%%%%%%%%%%%%%%%%%%%%%%%%%%%%%%%%%%%%%%%%%%%%%%%
\section{Threats to Validity}
\label{sec:threats}

Our experiments rely on randomized algorithms.
To take their randomness into account, experiments were repeated 10 times.

In our empirical study, 52 distinct APIs were used (9 toy examples, 8 examples for security research, and 36 from the WFD corpus).
Considering the use of 52 APIs, although this study might be currently the largest study in literature so far on open-source REST APIs,
a larger case study would be needed to be able to generalize our results, especially for APIs developed in industry.

Our novel oracles are not specific nor tailored to any fuzzer.
In our experiments, we used \evo, but in theory any other state-of-the-art fuzzer could had been used.
How easy or difficult would it be for our novel techniques to be integrated in other fuzzers is a matter that would require further investigation.

%%%%%%%%%%%%%%%%%%%%%%%%%%%%%%%%%%%%%%%%%%%%%%%%%%%%%%%%%%%%%%%%%%%%%%%%%%%%%%%%%%%%%%%%%%%%%%%%%%%%%%%%%%%%%%%%%%%%%%%
\section{Conclusion}
\label{sec:conclusions}

In this paper, we have presented nine novel oracles aimed at detecting authorization-related faults and common injection vulnerabilities in REST APIs.
These techniques were used to detect hundreds of new faults in an empirical study consisting of 52 REST APIs.

Our novel techniques were implemented as an extension of \evo, a state-of-the-art open-source fuzzer.
Once our novel techniques detect new faults, executable test cases (e.g., in JUnit format) are automatically generated to help practitioners in industry to reproduce the faults and debug them.

Future work will aim at designing new automated oracles to identify other kinds of security faults in REST APIs, and carry out empirical
studies in industry (for example at Fortune 500 enterprises that use \evo, like Volkswagen~\cite{icst2025vw} and Meituan~\cite{zhang2025fuzzing}).

Our extension to \evo is released as open-source at \texttt{\url{www.evomaster.org}}.
The novel techniques presented in this paper are now activated by default in \evo latest's releases.
The hundreds/thousands of practitioners in industry that use \evo  are already directly benefitting from the research work presented in this paper.

%%%%%%%%%%%%%%%%%%%%%%%%%%%%%%%%%%%%%%%%%%%%%%%%%%%%%%%%%%%%%%%%%%%%%%%%%%%%%%%%%%%%%%%%%%%%%%%%%%%%%%%%%%%%%%%%%%%%%%%
\section*{Data Availability}

All the techniques presented in this paper are implemented as part of \evo.
The fuzzer \evo is open-source on GitHub,\footnote{\url{https://github.com/WebFuzzing/EvoMaster}}
where each new release is automatically published on Zenodo for long-term storage (e.g.,~\cite{zenodo502evomaster}).

WFD is open-source on GitHub,\footnote{\url{https://github.com/WebFuzzing/Dataset}}
with as well each new release automatically uploaded to Zenodo for long-term storage (e.g.,~\cite{zenodo410wfd}).

The vulnerable APIs used for {\bf RQ2} are all open-source on GitHub, with IDs as specified in Table~\ref{tab:suts-bb}.
% and have been gathered in a new repository on GitHub to simplify replication.\footnote{https://github.com/WebFuzzing/security-apis}
% This repository also includes all needed files to run our experiments on these APIs, including authentication information files in WFC format.

%%%%%%%%%%%%%%%%%%%%%%%%%%%%%%%%%%%%%%%%%%%%%%%%%%%%%%%%%%%%%%%%%%%%%%%%%%%%%%%%%%%%%%%%%%%%%%%%%%%%%%%%%%%%%%%%%%%%%%%
\section*{Acknowledgment}

This work is funded by the European Research Council (ERC) under the European Union's Horizon 2020 research and innovation programme (EAST project, grant agreement No. 864972).
Omur Sahin is supported by the TÜBİTAK 2219 International Postdoctoral Research Fellowship Program (Project ID: 1059B192300060).
Man Zhang is supported by the National Science Foundation of China (grant agreement No. 62502022).

%%%%%%%%%%%%%%%%%%%%%%%%%%%%%%%%%%%%%%%%%%%%%%%%%%%%%%%%%%%%%%%%%%%%%%%%%%%%

\bibliographystyle{ACM-Reference-Format} % this requires ACM-Reference-Format.bst in same folder
%\bibliographystyle{acm}  % this is ancient from 80s, which does not support URL and DOI

%%https://arxiv.org/help/submit_tex#latex
%
% IMPORTANT: For final version arXiv, use generated bbl.
%            In such case, do not use compile.sh to build final pdf, but rather call pdflatex directly, eg
%            pdflatex arxiv


%\bibliography{../../../papers}

\begin{thebibliography}{51}

%%% the Web address.

\ifx \showCODEN    \undefined \def \showCODEN     #1{\unskip}     \fi
\ifx \showDOI      \undefined \def \showDOI       #1{#1}\fi
\ifx \showISBNx    \undefined \def \showISBNx     #1{\unskip}     \fi
\ifx \showISBNxiii \undefined \def \showISBNxiii  #1{\unskip}     \fi
\ifx \showISSN     \undefined \def \showISSN      #1{\unskip}     \fi
\ifx \showLCCN     \undefined \def \showLCCN      #1{\unskip}     \fi
\ifx \shownote     \undefined \def \shownote      #1{#1}          \fi
\ifx \showarticletitle \undefined \def \showarticletitle #1{#1}   \fi
\ifx \showURL      \undefined \def \showURL       {\relax}        \fi
% invisible to TeX
\providecommand\bibfield[2]{#2}
\providecommand\bibinfo[2]{#2}
\providecommand\natexlab[1]{#1}
\providecommand\showeprint[2][]{arXiv:#2}

\bibitem[\protect\citeauthoryear{Appelt, Nguyen, Briand, and Alshahwan}{Appelt
  et~al\mbox{.}}{2014}]%
        {appelt2014automated}
\bibfield{author}{\bibinfo{person}{Dennis Appelt}, \bibinfo{person}{Cu~Duy
  Nguyen}, \bibinfo{person}{Lionel~C Briand}, {and} \bibinfo{person}{Nadia
  Alshahwan}.} \bibinfo{year}{2014}\natexlab{}.
\newblock \showarticletitle{Automated testing for SQL injection
  vulnerabilities: an input mutation approach}. In
  \bibinfo{booktitle}{\emph{Proceedings of the 2014 International Symposium on
  Software Testing and Analysis}}. \bibinfo{pages}{259--269}.
\newblock


\bibitem[\protect\citeauthoryear{Arcuri}{Arcuri}{2017}]%
        {arcuri2017restful}
\bibfield{author}{\bibinfo{person}{Andrea Arcuri}.}
  \bibinfo{year}{2017}\natexlab{}.
\newblock \showarticletitle{{RESTful API Automated Test Case Generation}}. In
  \bibinfo{booktitle}{\emph{IEEE International Conference on Software Quality,
  Reliability and Security (QRS)}}. IEEE, \bibinfo{pages}{9--20}.
\newblock


\bibitem[\protect\citeauthoryear{Arcuri}{Arcuri}{2019}]%
        {arcuri2019restful}
\bibfield{author}{\bibinfo{person}{Andrea Arcuri}.}
  \bibinfo{year}{2019}\natexlab{}.
\newblock \showarticletitle{{RESTful API Automated Test Case Generation with
  EvoMaster}}.
\newblock \bibinfo{journal}{\emph{ACM Transactions on Software Engineering and
  Methodology (TOSEM)}} \bibinfo{volume}{28}, \bibinfo{number}{1}
  (\bibinfo{year}{2019}), \bibinfo{pages}{3}.
\newblock


\bibitem[\protect\citeauthoryear{Arcuri and Galeotti}{Arcuri and
  Galeotti}{2020}]%
        {arcuri2020sql}
\bibfield{author}{\bibinfo{person}{Andrea Arcuri} {and} \bibinfo{person}{Juan~P
  Galeotti}.} \bibinfo{year}{2020}\natexlab{}.
\newblock \showarticletitle{{Handling SQL databases in automated system test
  generation}}.
\newblock \bibinfo{journal}{\emph{ACM Transactions on Software Engineering and
  Methodology (TOSEM)}} \bibinfo{volume}{29}, \bibinfo{number}{4}
  (\bibinfo{year}{2020}), \bibinfo{pages}{1--31}.
\newblock


\bibitem[\protect\citeauthoryear{Arcuri, Poth, and Rrjolli}{Arcuri
  et~al\mbox{.}}{2025a}]%
        {icst2025vw}
\bibfield{author}{\bibinfo{person}{A. Arcuri}, \bibinfo{person}{A. Poth}, {and}
  \bibinfo{person}{O. Rrjolli}.} \bibinfo{year}{2025}\natexlab{a}.
\newblock \showarticletitle{Introducing Black-Box Fuzz Testing for REST APIs in
  Industry: Challenges and Solutions}. In \bibinfo{booktitle}{\emph{IEEE
  International Conference on Software Testing, Verification and Validation
  (ICST)}}.
\newblock


\bibitem[\protect\citeauthoryear{Arcuri, Sahin, and Zhang}{Arcuri
  et~al\mbox{.}}{2025b}]%
        {arcuri2025fuzzing}
\bibfield{author}{\bibinfo{person}{Andrea Arcuri}, \bibinfo{person}{Omur
  Sahin}, {and} \bibinfo{person}{Man Zhang}.} \bibinfo{year}{2025}\natexlab{b}.
\newblock \showarticletitle{Fuzzing for Detecting Access Policy Violations in
  REST APIs}. In \bibinfo{booktitle}{\emph{IEEE International Symposium on
  Software Reliability Engineering (ISSRE)}}.
\newblock


\bibitem[\protect\citeauthoryear{Arcuri, Zhang, Belhadi, Marculescu,
  Golmohammadi, Galeotti, and Seran}{Arcuri et~al\mbox{.}}{2023a}]%
        {arcuri2023building}
\bibfield{author}{\bibinfo{person}{Andrea Arcuri}, \bibinfo{person}{Man Zhang},
  \bibinfo{person}{Asma Belhadi}, \bibinfo{person}{Bogdan Marculescu},
  \bibinfo{person}{Amid Golmohammadi}, \bibinfo{person}{Juan~Pablo Galeotti},
  {and} \bibinfo{person}{Susruthan Seran}.} \bibinfo{year}{2023}\natexlab{a}.
\newblock \showarticletitle{Building an open-source system test generation
  tool: lessons learned and empirical analyses with EvoMaster}.
\newblock \bibinfo{journal}{\emph{Software Quality Journal}}
  (\bibinfo{year}{2023}), \bibinfo{pages}{1--44}.
\newblock


\bibitem[\protect\citeauthoryear{Arcuri, Zhang, Golmohammadi, Belhadi,
  Galeotti, Marculescu, and Seran}{Arcuri et~al\mbox{.}}{2023b}]%
        {icst2023emb}
\bibfield{author}{\bibinfo{person}{Andrea Arcuri}, \bibinfo{person}{Man Zhang},
  \bibinfo{person}{Amid Golmohammadi}, \bibinfo{person}{Asma Belhadi},
  \bibinfo{person}{Juan~P Galeotti}, \bibinfo{person}{Bogdan Marculescu}, {and}
  \bibinfo{person}{Susruthan Seran}.} \bibinfo{year}{2023}\natexlab{b}.
\newblock \showarticletitle{{EMB: A curated corpus of web/enterprise
  applications and library support for software testing research}}. In
  \bibinfo{booktitle}{\emph{2023 IEEE Conference on Software Testing,
  Verification and Validation (ICST)}}. IEEE, \bibinfo{pages}{433--442}.
\newblock


\bibitem[\protect\citeauthoryear{Arcuri, Zhang, Seran, Belhadi, Galeotti,
  Bogdan, Golmohammadi, Duman, Philip, Aldasoro, Ömür Şahin, Shalmani,
  Castagna, López, Ghianni, aszyrej, Rodriguez, Castagna, Roca, IvaK,
  Panichella, Niemeyer, and Maugeri}{Arcuri et~al\mbox{.}}{2026a}]%
        {zenodo502evomaster}
\bibfield{author}{\bibinfo{person}{Andrea Arcuri}, \bibinfo{person}{Man Zhang},
  \bibinfo{person}{Susruthan Seran}, \bibinfo{person}{Asma Belhadi},
  \bibinfo{person}{Juan~Pablo Galeotti}, \bibinfo{person}{Bogdan},
  \bibinfo{person}{Amid Golmohammadi}, \bibinfo{person}{Onur Duman},
  \bibinfo{person}{Philip}, \bibinfo{person}{Agustina Aldasoro},
  \bibinfo{person}{Ömür Şahin}, \bibinfo{person}{Mohsen~Taheri Shalmani},
  \bibinfo{person}{Franco Castagna}, \bibinfo{person}{Alberto~Martín López},
  \bibinfo{person}{Hernan Ghianni}, \bibinfo{person}{aszyrej},
  \bibinfo{person}{Miguel Rodriguez}, \bibinfo{person}{Franco~Nicolás
  Castagna}, \bibinfo{person}{Lucas~Mas Roca}, \bibinfo{person}{IvaK},
  \bibinfo{person}{Annibale Panichella}, \bibinfo{person}{Kyle Niemeyer}, {and}
  \bibinfo{person}{Marcello Maugeri}.} \bibinfo{year}{2026}\natexlab{a}.
\newblock \bibinfo{booktitle}{\emph{WebFuzzing/EvoMaster: v5.0.2}}.
\newblock
\urldef\tempurl%
\url{https://doi.org/10.5281/zenodo.18262322}
\showDOI{\tempurl}


\bibitem[\protect\citeauthoryear{Arcuri, Zhang, Seran, Galeotti, Golmohammadi,
  Duman, Aldasoro, and Ghianni}{Arcuri et~al\mbox{.}}{2025c}]%
        {arcuri2025tool}
\bibfield{author}{\bibinfo{person}{Andrea Arcuri}, \bibinfo{person}{Man Zhang},
  \bibinfo{person}{Susruthan Seran}, \bibinfo{person}{Juan~Pablo Galeotti},
  \bibinfo{person}{Amid Golmohammadi}, \bibinfo{person}{Onur Duman},
  \bibinfo{person}{Agustina Aldasoro}, {and} \bibinfo{person}{Hernan Ghianni}.}
  \bibinfo{year}{2025}\natexlab{c}.
\newblock \showarticletitle{Tool report: EvoMaster—black and white box
  search-based fuzzing for REST, GraphQL and RPC APIs}.
\newblock \bibinfo{journal}{\emph{Automated Software Engineering}}
  \bibinfo{volume}{32}, \bibinfo{number}{1} (\bibinfo{year}{2025}),
  \bibinfo{pages}{1--11}.
\newblock


\bibitem[\protect\citeauthoryear{Arcuri, Zhang, Seran, Galeotti, Golmohammadi,
  Garrett, Sahin, Shalmani, Castagna, Kertusha, Susilo, Bhandari, Assres, and
  Molléri}{Arcuri et~al\mbox{.}}{2026b}]%
        {arcuri2026sbft}
\bibfield{author}{\bibinfo{person}{Andrea Arcuri}, \bibinfo{person}{Man Zhang},
  \bibinfo{person}{Susruthan Seran}, \bibinfo{person}{Juan~P. Galeotti},
  \bibinfo{person}{Amid Golmohammadi}, \bibinfo{person}{Philip Garrett},
  \bibinfo{person}{Omur Sahin}, \bibinfo{person}{Mohsen~Taher Shalmani},
  \bibinfo{person}{Franco Castagna}, \bibinfo{person}{Iva Kertusha},
  \bibinfo{person}{Fanny~Febriani Susilo}, \bibinfo{person}{Guru~Prasad
  Bhandari}, \bibinfo{person}{Gebremariam Assres}, {and}
  \bibinfo{person}{Jefferson~Seide Molléri}.}
  \bibinfo{year}{2026}\natexlab{b}.
\newblock \showarticletitle{{EvoMaster at REST League 2026 Tool Competition}}.
  In \bibinfo{booktitle}{\emph{19th Search-Based and Fuzz Testing Workshop
  ({SBFT '26})}}.
\newblock


\bibitem[\protect\citeauthoryear{Arcuri, Zhang, Ömür Şahin, Golmohammadi,
  Belhadi, Seran, Duman, Galeotti, Assres, and Ghianni}{Arcuri
  et~al\mbox{.}}{2026c}]%
        {zenodo410wfd}
\bibfield{author}{\bibinfo{person}{Andrea Arcuri}, \bibinfo{person}{Man Zhang},
  \bibinfo{person}{Ömür Şahin}, \bibinfo{person}{Amid Golmohammadi},
  \bibinfo{person}{Asma Belhadi}, \bibinfo{person}{Susruthan Seran},
  \bibinfo{person}{Onur Duman}, \bibinfo{person}{Juan~Pablo Galeotti},
  \bibinfo{person}{Gebremariam~Mesfin Assres}, {and} \bibinfo{person}{Hernan
  Ghianni}.} \bibinfo{year}{2026}\natexlab{c}.
\newblock \bibinfo{booktitle}{\emph{WebFuzzing/Dataset: v4.1.0}}.
\newblock
\urldef\tempurl%
\url{https://doi.org/10.5281/zenodo.18196038}
\showDOI{\tempurl}


\bibitem[\protect\citeauthoryear{Atlidakis, Godefroid, and
  Polishchuk}{Atlidakis et~al\mbox{.}}{2019}]%
        {restlerICSE2019}
\bibfield{author}{\bibinfo{person}{Vaggelis Atlidakis},
  \bibinfo{person}{Patrice Godefroid}, {and} \bibinfo{person}{Marina
  Polishchuk}.} \bibinfo{year}{2019}\natexlab{}.
\newblock \showarticletitle{RESTler: Stateful {REST} {API} Fuzzing}. In
  \bibinfo{booktitle}{\emph{ACM/IEEE International Conference on Software
  Engineering (ICSE)}}. \bibinfo{pages}{748–758}.
\newblock


\bibitem[\protect\citeauthoryear{Atlidakis, Godefroid, and
  Polishchuk}{Atlidakis et~al\mbox{.}}{2020}]%
        {atlidakis2020checking}
\bibfield{author}{\bibinfo{person}{Vaggelis Atlidakis},
  \bibinfo{person}{Patrice Godefroid}, {and} \bibinfo{person}{Marina
  Polishchuk}.} \bibinfo{year}{2020}\natexlab{}.
\newblock \showarticletitle{Checking security properties of cloud service rest
  apis}. In \bibinfo{booktitle}{\emph{IEEE International Conference on Software
  Testing, Verification and Validation (ICST)}}. IEEE,
  \bibinfo{pages}{387--397}.
\newblock


\bibitem[\protect\citeauthoryear{Aziz, Bader, and Hippolyte}{Aziz
  et~al\mbox{.}}{2016}]%
        {aziz2016search}
\bibfield{author}{\bibinfo{person}{Benjamin Aziz}, \bibinfo{person}{Mohamed
  Bader}, {and} \bibinfo{person}{Cerana Hippolyte}.}
  \bibinfo{year}{2016}\natexlab{}.
\newblock \showarticletitle{Search-based sql injection attacks testing using
  genetic programming}. In \bibinfo{booktitle}{\emph{Genetic Programming: 19th
  European Conference, EuroGP 2016, Porto, Portugal, March 30-April 1, 2016,
  Proceedings 19}}. Springer, \bibinfo{pages}{183--198}.
\newblock


\bibitem[\protect\citeauthoryear{Barr, Harman, McMinn, Shahbaz, and Yoo}{Barr
  et~al\mbox{.}}{2015}]%
        {barr2015oracle}
\bibfield{author}{\bibinfo{person}{Earl~T Barr}, \bibinfo{person}{Mark Harman},
  \bibinfo{person}{Phil McMinn}, \bibinfo{person}{Muzammil Shahbaz}, {and}
  \bibinfo{person}{Shin Yoo}.} \bibinfo{year}{2015}\natexlab{}.
\newblock \showarticletitle{The oracle problem in software testing: A survey}.
\newblock \bibinfo{journal}{\emph{IEEE Transactions on Software Engineering
  (TSE)}} \bibinfo{volume}{41}, \bibinfo{number}{5} (\bibinfo{year}{2015}),
  \bibinfo{pages}{507--525}.
\newblock


\bibitem[\protect\citeauthoryear{Corradini, Montolli, Pasqua, and
  Ceccato}{Corradini et~al\mbox{.}}{2024}]%
        {corradini2024deeprest}
\bibfield{author}{\bibinfo{person}{Davide Corradini}, \bibinfo{person}{Zeno
  Montolli}, \bibinfo{person}{Michele Pasqua}, {and} \bibinfo{person}{Mariano
  Ceccato}.} \bibinfo{year}{2024}\natexlab{}.
\newblock \showarticletitle{DeepREST: Automated Test Case Generation for REST
  APIs Exploiting Deep Reinforcement Learning}. In
  \bibinfo{booktitle}{\emph{Proceedings of the 39th IEEE/ACM International
  Conference on Automated Software Engineering}}. \bibinfo{pages}{1383--1394}.
\newblock


\bibitem[\protect\citeauthoryear{Corradini, Pasqua, and Ceccato}{Corradini
  et~al\mbox{.}}{2023}]%
        {corradini2023automated}
\bibfield{author}{\bibinfo{person}{Davide Corradini}, \bibinfo{person}{Michele
  Pasqua}, {and} \bibinfo{person}{Mariano Ceccato}.}
  \bibinfo{year}{2023}\natexlab{}.
\newblock \showarticletitle{Automated black-box testing of mass assignment
  vulnerabilities in RESTful APIs}. In \bibinfo{booktitle}{\emph{2023 IEEE/ACM
  45th International Conference on Software Engineering (ICSE)}}. IEEE,
  \bibinfo{pages}{2553--2564}.
\newblock


\bibitem[\protect\citeauthoryear{Deng, Zhang, Li, Liu, Zhang, Liu, Yu, and
  Wang}{Deng et~al\mbox{.}}{2023}]%
        {deng2023nautilus}
\bibfield{author}{\bibinfo{person}{Gelei Deng}, \bibinfo{person}{Zhiyi Zhang},
  \bibinfo{person}{Yuekang Li}, \bibinfo{person}{Yi Liu},
  \bibinfo{person}{Tianwei Zhang}, \bibinfo{person}{Yang Liu},
  \bibinfo{person}{Guo Yu}, {and} \bibinfo{person}{Dongjin Wang}.}
  \bibinfo{year}{2023}\natexlab{}.
\newblock \showarticletitle{$\{$NAUTILUS$\}$: Automated
  $\{$RESTful$\}$$\{$API$\}$ Vulnerability Detection}. In
  \bibinfo{booktitle}{\emph{32nd USENIX Security Symposium (USENIX Security
  23)}}. \bibinfo{pages}{5593--5609}.
\newblock


\bibitem[\protect\citeauthoryear{Du, Li, Wang, Chen, Zhao, Zhu, Han, Wang, and
  Xue}{Du et~al\mbox{.}}{2024}]%
        {du2024vulnerability}
\bibfield{author}{\bibinfo{person}{Wenlong Du}, \bibinfo{person}{Jian Li},
  \bibinfo{person}{Yanhao Wang}, \bibinfo{person}{Libo Chen},
  \bibinfo{person}{Ruijie Zhao}, \bibinfo{person}{Junmin Zhu},
  \bibinfo{person}{Zhengguang Han}, \bibinfo{person}{Yijun Wang}, {and}
  \bibinfo{person}{Zhi Xue}.} \bibinfo{year}{2024}\natexlab{}.
\newblock \showarticletitle{Vulnerability-oriented testing for restful apis}.
  In \bibinfo{booktitle}{\emph{33rd USENIX Security Symposium (USENIX Security
  24)}}. USENIX Association, \bibinfo{pages}{739--755}.
\newblock


\bibitem[\protect\citeauthoryear{Foley and Maffeis}{Foley and Maffeis}{2025}]%
        {foley2025apirl}
\bibfield{author}{\bibinfo{person}{Myles Foley} {and} \bibinfo{person}{Sergio
  Maffeis}.} \bibinfo{year}{2025}\natexlab{}.
\newblock \showarticletitle{APIRL: Deep Reinforcement Learning for REST API
  Fuzzing}. In \bibinfo{booktitle}{\emph{Thirty-ninth Conference on Artificial
  Intelligence (AAAI 2025)}}.
\newblock


\bibitem[\protect\citeauthoryear{Golmohammadi, Zhang, and Arcuri}{Golmohammadi
  et~al\mbox{.}}{2023}]%
        {golmohammadi2023testing}
\bibfield{author}{\bibinfo{person}{Amid Golmohammadi}, \bibinfo{person}{Man
  Zhang}, {and} \bibinfo{person}{Andrea Arcuri}.}
  \bibinfo{year}{2023}\natexlab{}.
\newblock \showarticletitle{Testing RESTful APIs: A Survey}.
\newblock \bibinfo{journal}{\emph{ACM Transactions on Software Engineering and
  Methodology}} (\bibinfo{date}{aug} \bibinfo{year}{2023}).
\newblock
\showISSN{1049-331X}
\urldef\tempurl%
\url{https://doi.org/10.1145/3617175}
\showDOI{\tempurl}


\bibitem[\protect\citeauthoryear{Hatfield-Dodds and Dygalo}{Hatfield-Dodds and
  Dygalo}{2022}]%
        {hatfield2022deriving}
\bibfield{author}{\bibinfo{person}{Zac Hatfield-Dodds} {and}
  \bibinfo{person}{Dmitry Dygalo}.} \bibinfo{year}{2022}\natexlab{}.
\newblock \showarticletitle{Deriving Semantics-Aware Fuzzers from Web API
  Schemas}. In \bibinfo{booktitle}{\emph{2022 IEEE/ACM 44th International
  Conference on Software Engineering: Companion Proceedings (ICSE-Companion)}}.
  IEEE, \bibinfo{pages}{345--346}.
\newblock


\bibitem[\protect\citeauthoryear{Hydara, Sultan, Zulzalil, and
  Admodisastro}{Hydara et~al\mbox{.}}{2015}]%
        {hydara2015current}
\bibfield{author}{\bibinfo{person}{Isatou Hydara}, \bibinfo{person}{Abu
  Bakar~Md Sultan}, \bibinfo{person}{Hazura Zulzalil}, {and}
  \bibinfo{person}{Novia Admodisastro}.} \bibinfo{year}{2015}\natexlab{}.
\newblock \showarticletitle{Current state of research on cross-site scripting
  (XSS)--A systematic literature review}.
\newblock \bibinfo{journal}{\emph{Information and Software Technology}}
  \bibinfo{volume}{58} (\bibinfo{year}{2015}), \bibinfo{pages}{170--186}.
\newblock


\bibitem[\protect\citeauthoryear{Kieyzun, Guo, Jayaraman, and Ernst}{Kieyzun
  et~al\mbox{.}}{2009}]%
        {KGJE09}
\bibfield{author}{\bibinfo{person}{A. Kieyzun}, \bibinfo{person}{P.J. Guo},
  \bibinfo{person}{K. Jayaraman}, {and} \bibinfo{person}{M.D. Ernst}.}
  \bibinfo{year}{2009}\natexlab{}.
\newblock \showarticletitle{{Automatic creation of SQL injection and cross-site
  scripting attacks}}. In \bibinfo{booktitle}{\emph{ACM/IEEE International
  Conference on Software Engineering (ICSE)}}. \bibinfo{pages}{199--209}.
\newblock


\bibitem[\protect\citeauthoryear{Kim, Sinha, and Orso}{Kim
  et~al\mbox{.}}{2023}]%
        {kim2023adaptive}
\bibfield{author}{\bibinfo{person}{Myeongsoo Kim}, \bibinfo{person}{Saurabh
  Sinha}, {and} \bibinfo{person}{Alessandro Orso}.}
  \bibinfo{year}{2023}\natexlab{}.
\newblock \showarticletitle{Adaptive rest api testing with reinforcement
  learning}. In \bibinfo{booktitle}{\emph{2023 38th IEEE/ACM International
  Conference on Automated Software Engineering (ASE)}}. IEEE,
  \bibinfo{pages}{446--458}.
\newblock


\bibitem[\protect\citeauthoryear{Kim, Sinha, and Orso}{Kim
  et~al\mbox{.}}{2025a}]%
        {kim2025llamaresttest}
\bibfield{author}{\bibinfo{person}{Myeongsoo Kim}, \bibinfo{person}{Saurabh
  Sinha}, {and} \bibinfo{person}{Alessandro Orso}.}
  \bibinfo{year}{2025}\natexlab{a}.
\newblock \showarticletitle{LlamaRestTest: Effective REST API Testing with
  Small Language Models}. In \bibinfo{booktitle}{\emph{ACM Symposium on the
  Foundations of Software Engineering (FSE)}}.
\newblock


\bibitem[\protect\citeauthoryear{Kim, Stennett, Sinha, and Orso}{Kim
  et~al\mbox{.}}{2025b}]%
        {kim2025autoresttest}
\bibfield{author}{\bibinfo{person}{Myeongsoo Kim}, \bibinfo{person}{Tyler
  Stennett}, \bibinfo{person}{Saurabh Sinha}, {and} \bibinfo{person}{Alessandro
  Orso}.} \bibinfo{year}{2025}\natexlab{b}.
\newblock \showarticletitle{A Multi-Agent Approach for REST API Testing with
  Semantic Graphs and LLM-Driven Inputs}.
\newblock \bibinfo{journal}{\emph{ACM/IEEE International Conference on Software
  Engineering (ICSE)}} (\bibinfo{year}{2025}).
\newblock


\bibitem[\protect\citeauthoryear{Kim, Xin, Sinha, and Orso}{Kim
  et~al\mbox{.}}{2022}]%
        {Kim2022Rest}
\bibfield{author}{\bibinfo{person}{Myeongsoo Kim}, \bibinfo{person}{Qi Xin},
  \bibinfo{person}{Saurabh Sinha}, {and} \bibinfo{person}{Alessandro Orso}.}
  \bibinfo{year}{2022}\natexlab{}.
\newblock \showarticletitle{{Automated Test Generation for REST APIs: No Time
  to Rest Yet}}. In \bibinfo{booktitle}{\emph{Proceedings of the 31st ACM
  SIGSOFT International Symposium on Software Testing and Analysis}}
  \emph{(\bibinfo{series}{ISSTA 2022})}. \bibinfo{publisher}{Association for
  Computing Machinery}, \bibinfo{address}{New York, NY, USA},
  \bibinfo{pages}{289–301}.
\newblock
\showISBNx{9781450393799}
\urldef\tempurl%
\url{https://doi.org/10.1145/3533767.3534401}
\showDOI{\tempurl}


\bibitem[\protect\citeauthoryear{Laranjeiro, Agnelo, and Bernardino}{Laranjeiro
  et~al\mbox{.}}{2021}]%
        {laranjeiro2021black}
\bibfield{author}{\bibinfo{person}{Nuno Laranjeiro}, \bibinfo{person}{Jo{\~a}o
  Agnelo}, {and} \bibinfo{person}{Jorge Bernardino}.}
  \bibinfo{year}{2021}\natexlab{}.
\newblock \showarticletitle{A black box tool for robustness testing of REST
  services}.
\newblock \bibinfo{journal}{\emph{IEEE Access}}  \bibinfo{volume}{9}
  (\bibinfo{year}{2021}), \bibinfo{pages}{24738--24754}.
\newblock


\bibitem[\protect\citeauthoryear{Liu, Li, and Chen}{Liu et~al\mbox{.}}{2020}]%
        {liu2020deepsqli}
\bibfield{author}{\bibinfo{person}{Muyang Liu}, \bibinfo{person}{Ke Li}, {and}
  \bibinfo{person}{Tao Chen}.} \bibinfo{year}{2020}\natexlab{}.
\newblock \showarticletitle{DeepSQLi: Deep semantic learning for testing SQL
  injection}. In \bibinfo{booktitle}{\emph{Proceedings of the 29th ACM SIGSOFT
  International Symposium on Software Testing and Analysis}}.
  \bibinfo{pages}{286--297}.
\newblock


\bibitem[\protect\citeauthoryear{Liu, Li, Deng, Liu, Wan, Wu, Ji, Xu, and
  Bao}{Liu et~al\mbox{.}}{2022}]%
        {liu2022icse}
\bibfield{author}{\bibinfo{person}{Yi Liu}, \bibinfo{person}{Yuekang Li},
  \bibinfo{person}{Gelei Deng}, \bibinfo{person}{Yang Liu},
  \bibinfo{person}{Ruiyuan Wan}, \bibinfo{person}{Runchao Wu},
  \bibinfo{person}{Dandan Ji}, \bibinfo{person}{Shiheng Xu}, {and}
  \bibinfo{person}{Minli Bao}.} \bibinfo{year}{2022}\natexlab{}.
\newblock \showarticletitle{Morest: Model-based RESTful API Testing with
  Execution Feedback}. In \bibinfo{booktitle}{\emph{ACM/IEEE International
  Conference on Software Engineering (ICSE)}}.
\newblock


\bibitem[\protect\citeauthoryear{Lyu, Xu, Ji, Zhang, Wang, Zhao, Pan, Cao,
  Chen, and Beyah}{Lyu et~al\mbox{.}}{2023}]%
        {lyu2023miner}
\bibfield{author}{\bibinfo{person}{Chenyang Lyu}, \bibinfo{person}{Jiacheng
  Xu}, \bibinfo{person}{Shouling Ji}, \bibinfo{person}{Xuhong Zhang},
  \bibinfo{person}{Qinying Wang}, \bibinfo{person}{Binbin Zhao},
  \bibinfo{person}{Gaoning Pan}, \bibinfo{person}{Wei Cao},
  \bibinfo{person}{Peng Chen}, {and} \bibinfo{person}{Raheem Beyah}.}
  \bibinfo{year}{2023}\natexlab{}.
\newblock \showarticletitle{MINER: A Hybrid Data-Driven Approach for REST API
  Fuzzing}. In \bibinfo{booktitle}{\emph{32nd USENIX Security Symposium (USENIX
  Security 23)}}. \bibinfo{pages}{4517--4534}.
\newblock


\bibitem[\protect\citeauthoryear{Marculescu, Zhang, and Arcuri}{Marculescu
  et~al\mbox{.}}{2022}]%
        {marculescu2022faults}
\bibfield{author}{\bibinfo{person}{Bogdan Marculescu}, \bibinfo{person}{Man
  Zhang}, {and} \bibinfo{person}{Andrea Arcuri}.}
  \bibinfo{year}{2022}\natexlab{}.
\newblock \showarticletitle{On the Faults Found in REST APIs by Automated Test
  Generation}.
\newblock \bibinfo{journal}{\emph{ACM Transactions on Software Engineering and
  Methodology (TOSEM)}} \bibinfo{volume}{31}, \bibinfo{number}{3}
  (\bibinfo{year}{2022}), \bibinfo{pages}{1--43}.
\newblock


\bibitem[\protect\citeauthoryear{Martin, Xie, and Yu}{Martin
  et~al\mbox{.}}{2006}]%
        {martin2006defining}
\bibfield{author}{\bibinfo{person}{Evan Martin}, \bibinfo{person}{Tao Xie},
  {and} \bibinfo{person}{Ting Yu}.} \bibinfo{year}{2006}\natexlab{}.
\newblock \showarticletitle{Defining and measuring policy coverage in testing
  access control policies}. In \bibinfo{booktitle}{\emph{Information and
  Communications Security: 8th International Conference, ICICS 2006, Raleigh,
  NC, USA, December 4-7, 2006. Proceedings 8}}. Springer,
  \bibinfo{pages}{139--158}.
\newblock


\bibitem[\protect\citeauthoryear{Martin-Lopez, Segura, and
  Ruiz-Cort{\'e}s}{Martin-Lopez et~al\mbox{.}}{2019}]%
        {martin2019test}
\bibfield{author}{\bibinfo{person}{Alberto Martin-Lopez},
  \bibinfo{person}{Sergio Segura}, {and} \bibinfo{person}{Antonio
  Ruiz-Cort{\'e}s}.} \bibinfo{year}{2019}\natexlab{}.
\newblock \showarticletitle{Test coverage criteria for RESTful web APIs}. In
  \bibinfo{booktitle}{\emph{Proceedings of the 10th ACM SIGSOFT International
  Workshop on Automating TEST Case Design, Selection, and Evaluation}}.
  \bibinfo{pages}{15--21}.
\newblock


\bibitem[\protect\citeauthoryear{Martin-Lopez, Segura, and
  Ruiz-Cort\'{e}s}{Martin-Lopez et~al\mbox{.}}{2021}]%
        {martinLopez2021Restest}
\bibfield{author}{\bibinfo{person}{Alberto Martin-Lopez},
  \bibinfo{person}{Sergio Segura}, {and} \bibinfo{person}{Antonio
  Ruiz-Cort\'{e}s}.} \bibinfo{year}{2021}\natexlab{}.
\newblock \showarticletitle{{RESTest: Automated Black-Box Testing of RESTful
  Web APIs}}. In \bibinfo{booktitle}{\emph{ACM Int. Symposium on Software
  Testing and Analysis (ISSTA)}}. \bibinfo{publisher}{ACM},
  \bibinfo{pages}{682--685}.
\newblock


\bibitem[\protect\citeauthoryear{Pasca, Delinschi, Erdei, and Matei}{Pasca
  et~al\mbox{.}}{2025}]%
        {pasca2025llm}
\bibfield{author}{\bibinfo{person}{Emil~Marian Pasca}, \bibinfo{person}{Daniela
  Delinschi}, \bibinfo{person}{Rudolf Erdei}, {and} \bibinfo{person}{Oliviu
  Matei}.} \bibinfo{year}{2025}\natexlab{}.
\newblock \showarticletitle{LLM-Driven, Self-Improving Framework for Security
  Test Automation: Leveraging Karate DSL for Augmented API Resilience}.
\newblock \bibinfo{journal}{\emph{IEEE Access}} (\bibinfo{year}{2025}).
\newblock


\bibitem[\protect\citeauthoryear{Poth, Rrjolli, and Arcuri}{Poth
  et~al\mbox{.}}{2025}]%
        {poth2025technology}
\bibfield{author}{\bibinfo{person}{Alexander Poth}, \bibinfo{person}{Olsi
  Rrjolli}, {and} \bibinfo{person}{Andrea Arcuri}.}
  \bibinfo{year}{2025}\natexlab{}.
\newblock \showarticletitle{Technology adoption performance evaluation applied
  to testing industrial REST APIs}.
\newblock \bibinfo{journal}{\emph{Automated Software Engineering}}
  \bibinfo{volume}{32}, \bibinfo{number}{1} (\bibinfo{year}{2025}),
  \bibinfo{pages}{5}.
\newblock


\bibitem[\protect\citeauthoryear{Rooijakkers, Nijsten, Daniele, Weitenberg,
  Groenewegen, and Melissen}{Rooijakkers et~al\mbox{.}}{2025}]%
        {rooijakkers2025wuppiefuzz}
\bibfield{author}{\bibinfo{person}{Thomas Rooijakkers}, \bibinfo{person}{Anne
  Nijsten}, \bibinfo{person}{Cristian Daniele}, \bibinfo{person}{Erieke
  Weitenberg}, \bibinfo{person}{Ringo Groenewegen}, {and}
  \bibinfo{person}{Arthur Melissen}.} \bibinfo{year}{2025}\natexlab{}.
\newblock \showarticletitle{WuppieFuzz: Coverage-Guided, Stateful REST API
  Fuzzing}.
\newblock \bibinfo{journal}{\emph{arXiv preprint arXiv:2512.15554}}
  (\bibinfo{year}{2025}).
\newblock


\bibitem[\protect\citeauthoryear{Sahin, Zhang, and Arcuri}{Sahin
  et~al\mbox{.}}{2025}]%
        {sahin_2025_wfc}
\bibfield{author}{\bibinfo{person}{Omur Sahin}, \bibinfo{person}{Man Zhang},
  {and} \bibinfo{person}{Andrea Arcuri}.} \bibinfo{year}{2025}\natexlab{}.
\newblock \bibinfo{title}{WFC/WFD: Web Fuzzing Commons, Dataset and Guidelines
  to Support Experimentation in REST API Fuzzing}.
\newblock
\newblock
\showeprint[arxiv]{cs.SE/2509.01612}
\urldef\tempurl%
\url{https://arxiv.org/abs/2509.01612}
\showURL{%
\tempurl}


\bibitem[\protect\citeauthoryear{Santos~Filho, Rodr{\'\i}guez, and
  Feitosa}{Santos~Filho et~al\mbox{.}}{2025}]%
        {santos2025automated}
\bibfield{author}{\bibinfo{person}{Ailton Santos~Filho},
  \bibinfo{person}{Ricardo~J Rodr{\'\i}guez}, {and} \bibinfo{person}{Eduardo~L
  Feitosa}.} \bibinfo{year}{2025}\natexlab{}.
\newblock \showarticletitle{Automated broken object-level authorization attack
  detection in REST APIs through OpenAPI to colored petri nets transformation}.
\newblock \bibinfo{journal}{\emph{International Journal of Information
  Security}} \bibinfo{volume}{24}, \bibinfo{number}{2} (\bibinfo{year}{2025}),
  \bibinfo{pages}{83}.
\newblock


\bibitem[\protect\citeauthoryear{Sartaj, Ali, and Gjøby}{Sartaj
  et~al\mbox{.}}{2024}]%
        {sartaj2024restapitestingdevops}
\bibfield{author}{\bibinfo{person}{Hassan Sartaj}, \bibinfo{person}{Shaukat
  Ali}, {and} \bibinfo{person}{Julie~Marie Gjøby}.}
  \bibinfo{year}{2024}\natexlab{}.
\newblock \bibinfo{title}{REST API Testing in DevOps: A Study on an Evolving
  Healthcare IoT Application}.
\newblock
\newblock
\showeprint[arxiv]{cs.SE/2410.12547}
\urldef\tempurl%
\url{https://arxiv.org/abs/2410.12547}
\showURL{%
\tempurl}


\bibitem[\protect\citeauthoryear{Seran, Bhandari, and Arcuri}{Seran
  et~al\mbox{.}}{2026}]%
        {seran2026sbft}
\bibfield{author}{\bibinfo{person}{Susruthan Seran},
  \bibinfo{person}{Guru~Prasad Bhandari}, {and} \bibinfo{person}{Andrea
  Arcuri}.} \bibinfo{year}{2026}\natexlab{}.
\newblock \showarticletitle{{Detecting Server-Side Request Forgery (SSRF)
  Vulnerabilities In REST API Fuzz Testing}}. In \bibinfo{booktitle}{\emph{19th
  Search-Based and Fuzz Testing Workshop ({SBFT '26})}}.
\newblock


\bibitem[\protect\citeauthoryear{Viglianisi, Dallago, and Ceccato}{Viglianisi
  et~al\mbox{.}}{2020}]%
        {viglianisi2020resttestgen}
\bibfield{author}{\bibinfo{person}{Emanuele Viglianisi},
  \bibinfo{person}{Michael Dallago}, {and} \bibinfo{person}{Mariano Ceccato}.}
  \bibinfo{year}{2020}\natexlab{}.
\newblock \showarticletitle{RESTTESTGEN: Automated Black-Box Testing of RESTful
  APIs}. In \bibinfo{booktitle}{\emph{IEEE International Conference on Software
  Testing, Verification and Validation (ICST)}}. IEEE.
\newblock


\bibitem[\protect\citeauthoryear{Wang and Xu}{Wang and Xu}{2024}]%
        {wang2024beyond}
\bibfield{author}{\bibinfo{person}{Yu Wang} {and} \bibinfo{person}{Yue Xu}.}
  \bibinfo{year}{2024}\natexlab{}.
\newblock \showarticletitle{Beyond REST: Introducing APIF for Comprehensive API
  Vulnerability Fuzzing}. In \bibinfo{booktitle}{\emph{Proceedings of the 27th
  International Symposium on Research in Attacks, Intrusions and Defenses}}.
  \bibinfo{pages}{435--449}.
\newblock


\bibitem[\protect\citeauthoryear{Wu, Xu, Niu, and Nie}{Wu
  et~al\mbox{.}}{2022}]%
        {wu2022icse}
\bibfield{author}{\bibinfo{person}{Huayao Wu}, \bibinfo{person}{Lixin Xu},
  \bibinfo{person}{Xintao Niu}, {and} \bibinfo{person}{Changhai Nie}.}
  \bibinfo{year}{2022}\natexlab{}.
\newblock \showarticletitle{Combinatorial Testing of RESTful APIs}. In
  \bibinfo{booktitle}{\emph{ACM/IEEE International Conference on Software
  Engineering (ICSE)}}.
\newblock


\bibitem[\protect\citeauthoryear{Zhang and Arcuri}{Zhang and Arcuri}{2023}]%
        {zhang2023open}
\bibfield{author}{\bibinfo{person}{Man Zhang} {and} \bibinfo{person}{Andrea
  Arcuri}.} \bibinfo{year}{2023}\natexlab{}.
\newblock \showarticletitle{{Open Problems in Fuzzing RESTful APIs: A
  Comparison of Tools}}.
\newblock \bibinfo{journal}{\emph{ACM Transactions on Software Engineering and
  Methodology (TOSEM)}} (\bibinfo{date}{may} \bibinfo{year}{2023}).
\newblock
\showISSN{1049-331X}
\urldef\tempurl%
\url{https://doi.org/10.1145/3597205}
\showDOI{\tempurl}


\bibitem[\protect\citeauthoryear{Zhang, Arcuri, Li, Liu, and Xue}{Zhang
  et~al\mbox{.}}{2023}]%
        {zhang2023rpc}
\bibfield{author}{\bibinfo{person}{Man Zhang}, \bibinfo{person}{Andrea Arcuri},
  \bibinfo{person}{Yonggang Li}, \bibinfo{person}{Yang Liu}, {and}
  \bibinfo{person}{Kaiming Xue}.} \bibinfo{year}{2023}\natexlab{}.
\newblock \showarticletitle{{White-Box Fuzzing RPC-Based APIs with EvoMaster:
  An Industrial Case Study}}.
\newblock \bibinfo{journal}{\emph{ACM Transactions on Software Engineering and
  Methodology}} \bibinfo{volume}{32}, \bibinfo{number}{5}
  (\bibinfo{year}{2023}), \bibinfo{pages}{1--38}.
\newblock


\bibitem[\protect\citeauthoryear{Zhang, Arcuri, Li, Liu, Xue, Wang, Huo, and
  Huang}{Zhang et~al\mbox{.}}{2025}]%
        {zhang2025fuzzing}
\bibfield{author}{\bibinfo{person}{Man Zhang}, \bibinfo{person}{Andrea Arcuri},
  \bibinfo{person}{Yonggang Li}, \bibinfo{person}{Yang Liu},
  \bibinfo{person}{Kaiming Xue}, \bibinfo{person}{Zhao Wang},
  \bibinfo{person}{Jian Huo}, {and} \bibinfo{person}{Weiwei Huang}.}
  \bibinfo{year}{2025}\natexlab{}.
\newblock \showarticletitle{Fuzzing microservices: A series of user studies in
  industry on industrial systems with evomaster}.
\newblock \bibinfo{journal}{\emph{Science of Computer Programming}}
  (\bibinfo{year}{2025}), \bibinfo{pages}{103322}.
\newblock


\bibitem[\protect\citeauthoryear{Zhang, Arcuri, Teng, Xue, and Wang}{Zhang
  et~al\mbox{.}}{2024}]%
        {zhang2024seeding}
\bibfield{author}{\bibinfo{person}{Man Zhang}, \bibinfo{person}{Andrea Arcuri},
  \bibinfo{person}{Piyun Teng}, \bibinfo{person}{Kaiming Xue}, {and}
  \bibinfo{person}{Wenhao Wang}.} \bibinfo{year}{2024}\natexlab{}.
\newblock \showarticletitle{Seeding and Mocking in White-Box Fuzzing Enterprise
  RPC APIs: An Industrial Case Study}. In \bibinfo{booktitle}{\emph{Proceedings
  of the 39th IEEE/ACM International Conference on Automated Software
  Engineering}}. \bibinfo{pages}{2024--2034}.
\newblock


\end{thebibliography}

%If needed
%\input{appendix.tex}

\end{document}